\documentclass[12pt]{article}
\usepackage{mathtools}
\usepackage{graphicx}
\usepackage[super,comma,sort&compress]{natbib}
\usepackage[margin=1in]{geometry}
\usepackage[english]{babel}
\usepackage{longtable}
\usepackage{color}
\usepackage{float}
\usepackage{amssymb,amsmath,amsthm}
\usepackage{multirow}
\usepackage[titletoc,title]{appendix}
\usepackage{authblk}
\usepackage{setspace}
\usepackage{dsfont}
\usepackage[OT1]{fontenc}
\usepackage{refcount}
\usepackage[colorlinks,citecolor=blue,urlcolor=blue]{hyperref}
\usepackage{float}
\usepackage{mathrsfs}
\usepackage{tabulary}
\usepackage{rotating}
\usepackage{setspace}
\usepackage{multirow}
\usepackage{paralist}
\usepackage{booktabs}
\usepackage{bm}
\usepackage{longtable}
\usepackage{bbm}
\usepackage{setspace}
\usepackage{array}
\usepackage{makecell}

\usepackage{utfsym}
\usepackage{makecell}
\usepackage{adjustbox}
\usepackage{listings}
\usepackage{lscape}
\usepackage[T1]{fontenc}
\usepackage[utf8]{inputenc}
\usepackage[english]{babel}
\usepackage{dsfont}
\usepackage{hyperref}
\usepackage[para]{threeparttable} 
\usepackage{quiver}

\renewcommand{\d}{\mathsf{d}}
\usepackage{url}
\usepackage{booktabs}
\usepackage{authblk}
\usepackage{etoolbox}
\usepackage{longtable}
\usepackage{caption}
\usepackage{hyperref}
\usepackage{nameref,zref-xr} 
\zxrsetup{toltxlabel} 
\usepackage{subfig}
\usepackage{listings}
\usepackage{xcolor}

\definecolor{codegreen}{rgb}{0,0.6,0}
\definecolor{codegray}{rgb}{0.5,0.5,0.5}
\definecolor{codepurple}{rgb}{0.58,0,0.82}
\definecolor{backcolour}{rgb}{0.95,0.95,0.92}

\lstdefinestyle{mystyle}{
    backgroundcolor=\color{backcolour},   
    commentstyle=\color{codegreen},
    numberstyle=\tiny\color{codegray},
    stringstyle=\color{codepurple},
    basicstyle=\ttfamily,
    breakatwhitespace=false,         
    breaklines=true,                 
    captionpos=b,                    
    keepspaces=true,                 
    numbers=left,                    
    numbersep=5pt,                  
    showspaces=false,                
    showstringspaces=false,
    showtabs=false,                  
    tabsize=2
}

\newtheorem{example}{Example}
\AtEndDocument{\refstepcounter{theorem}\label{finalthm}}
\AtEndDocument{\refstepcounter{equation}\label{finaleq}}
\usepackage{algorithm}
\usepackage{algpseudocode}

{
  \theoremstyle{definition}
  \newtheorem{assumption}{}
}
{
  \theoremstyle{definition}
  
}

\AtEndDocument{\refstepcounter{lemma}\label{finallemma}}

\renewcommand{\P}{\mathsf{P}}

\newcommand{\E}{\mathsf{E}}

\DeclarePairedDelimiterX{\norm}[1]{\lVert}{\rVert}{#1}

\pgfdeclarelayer{background}
\pgfsetlayers{background,main}

\date{\today}
\author[1]{Katherine L. Hoffman\thanks{Corresponding author: \\
kh3233@cumc.columbia.edu \\ 722 W. 168th St, NY, NY 10032}}
\author[2]{Diego Salazar-Barreto}
\author[3]{Nicholas T. Williams}
\author[3]{Kara E. Rudolph\thanks{Authors contributed equally.}}
\author[4]{Iv\'an D\'iaz$^\dagger$}

\affil[1]{\small Division of Biostatistics, Department of Population Health Science, Weill Cornell Medicine, New York, NY, USA.}
\affil[2]{\small  School of Industrial Engineering, University of Los Andes, Bogotá, Colombia.}
\affil[3]{\small Department of Epidemiology, Mailman School of Public Health, Columbia University, New York, NY, USA.}
\affil[4]{\small  Division of Biostatistics, Department of Population Health, New York University Grossman School of Medicine, New York University, New York, NY, USA.}

\begin{document}









\begin{abstract}
This tutorial discusses methodology for causal inference using \textit{longitudinal modified treatment policies}
(LMTPs). LMTPs facilitate the
mathematical formalization, identification, and estimation of many
novel parameters, and mathematically generalize many commonly used parameters, such as the average treatment effect. LMTPs apply to 
a wide variety of exposures, including binary, multivariate, and
continuous, and can accommodate time-varying treatments
and confounders, competing risks, loss-to-follow-up, as well as
survival, binary, or continuous
outcomes. LMTPs can be seen as an extension of static and dynamic interventions to involve the \emph{natural value of treatment}, and, like dynamic interventions, can be used to define alternative estimands with a positivity assumption that is more likely to be satisfied than estimands corresponding to static interventions. This tutorial aims to illustrate several practical uses of the LMTP methodology, including describing different estimation
strategies and their corresponding advantages and disadvantages. We provide numerous examples of types of research questions which can
be answered using LMTPs. We go into more depth with one of these examples---specifically, estimating the effect of delaying
intubation on critically ill COVID-19 patients'
mortality. We demonstrate the use of the open-source R package \emph{lmtp} to estimate the effects, and we provide code on 
\href{https://github.com/kathoffman/lmtp-tutorial}{https://github.com/kathoffman/lmtp-tutorial}.
\end{abstract}


\newpage


\section{Introduction}


Consider the research question, ``does early initiation of invasive mechanical ventilation (intubation) for coronavirus 2019 (COVID-19) patients increase mortality rates?". This query is the subject of active critical care research \citep{KRISHNAN2022154045, hyman2020timing, lee2020clinical, mckay2022comparison, ridjab2022outcome, papoutsi2021effect, hernandez2020timing, matta2020timing, pandya2021ventilatory, mellado2021high, bavishi2021timing}, but it is difficult to translate into a meaningful estimand and estimation procedure 
for multiple reasons. First, intubation can take place throughout a patient's course of hospitalization, meaning it is a time-varying exposure, or treatment, of interest. Time-varying exposures require specific methodology from the causal inference literature to 
properly accomodate time-dependent confounding \citep{mansournia2017handling, hernan2023causal}.
Second, respiratory support is a multilevel exposure, minimally consisting of categories such as ``no oxygen support,'' ``non-invasive mechanical ventilation,'' and ``invasive mechanical ventilation.'' The majority of the causal inference literature focuses on dichotomous exposures and interventions, e.g. treat everyone vs. treat no one, whereas methodology which can incorporate multilevel or continuous exposures is limited. Finally, a key assumption for causal inference is positivity, which roughly states that the intervention considered must occur with positive probability within all strata of confounders. For time-varying exposures, this positive probability must hold at each time point \citep{petersen2012diagnosing}. 


Longitudinal modified treatment policies (LMTPs) may offer a solution to the aforementioned challenges. LMTPs provide an
approach to translating complex research questions into 
a broad range of causal estimands, identifying such quantities, and estimating them. 
In brief, a modified treatment policy (MTP)
allows a
hypothetical intervention to depend on the \textit{natural value of
  treatment}, i.e. the value that treatment would take at time $t$ if an intervention was discontinued right before time $t$  \citep{YoungHernanRobins2014}. An MTP for a time-varying, or longitudinal, exposure, is an LMTP. The methodology for LMTPs not only mathematically generalizes MTPs, but can also be seen as an extension of more commonly known interventions such as \textit{static} and \textit{dynamic} interventions. The LMTP methodology thus allows for a wide range of
interventions \citep{robins2004effects, taubman2009intervening, shpitser2012effects, munoz2012population,  haneuse2013estimation,
richardson2013single, YoungHernanRobins2014, diaz_nonparametric_2021} including those on binary,
categorical, continuous, and multiple exposures. LMTP can also accommodate many types of outcomes including binary, continuous, or time-to-event outcomes with possible
competing risks and informative right-censoring. 
Finally, LMTPs can address violations of the positivity
assumption because they allow researchers to define, identify, and estimate alternative estimands for which positivity holds by design.

In this tutorial, we provide a guide to understanding and applying the LMTP methodology. We begin by reviewing static and dynamic interventions to see how LMTPs fit into the broader causal inference literature, and to understand how they mathematically generalize these interventions, as well as some other interventions which depend on the natural value of treatment. We then discuss specifics of the LMTP methodology, including defining interventions and identifying estimands. We highlight
several estimation procedures 
and provide numerous
examples of research questions that can be addressed with LMTPs. Lastly, we demonstrate an analysis on a real-world longitudinal observational data set. One way to operationalize the initially proposed question of the effect of intubation timing policies on mortality is to consider early versus late intubation, relative to the actual time patients were intubated. 
We illustrate one way of defining that query using an LMTP and provide
detailed descriptions of the study design and analytical methods, as
well as code and synthetic data to aid
future researchers.
 
\section{Notation and general setup}

Consider a sample of i.i.d. observations $Z_1, ..., Z_n$ drawn from a
distribution $\P$. This $\P$ represents a longitudinal process and may
contain any number of time points, but for simplicity we will describe a distribution
with only two time points, $t \in \{0,1\}$. For each unit in the
study, we observe a set of random variables
$Z = (L_0, A_0, L_1, A_1, Y)$. At the first time point, baseline
covariates $L_0$ affect the baseline exposure, $A_0$. At the second
time point, we observe covariates $L_1$ and exposure $A_1$, which are
themselves affected by $L_0$ and $A_0$, and have the potential to
change from their respective baseline values (time-varying). An
outcome $Y$ is measured at the end of a defined follow up period. Each
endogenous variable $L_0, A_0, L_1, A_1,$ and $Y$ has a corresponding
exogenous variable $U$, representing the unmeasured, external factors
affecting each measured process. We may use the following simplified directed acyclic graph (DAG) \citep{pearl1998graphs} to denote the set-up.

\[\begin{tikzcd}[row sep=3.15em]
	& {A_0} && {A_1} \\
	{L_0} && {L_1} && Y
	\arrow[from=2-1, to=2-3]
	\arrow[from=1-2, to=2-3]
	\arrow[from=2-3, to=1-4]
	\arrow[from=1-4, to=2-5]
	\arrow[from=2-3, to=2-5]
	\arrow[from=1-2, to=1-4]
	\arrow[from=2-1, to=1-2]
	\arrow[from=1-2, to=2-5]
	\arrow[curve={height=24pt}, from=2-1, to=2-5]
\end{tikzcd}\]

We will use $H_t$ as a shorthand notation for the history of data
measured up to right before $A_t$. For example, $H_0=L_0$, and
$H_1 = (L_0, A_0, L_1)$. We conceptualize causal interventions, or treatment policies, in
terms of hypothetical interventions on nodes of the DAG
\citep{pearl2016causal}.
First, consider a user-given function $\d_0(a_0, h_0, \epsilon_0)$ which maps a treatment value $a_0$, a history $h_0$, and a randomizer $\epsilon_0$ into a potential treatment value. 
The intervention at time $t=0$ is defined by removing node $A_0$ from the DAG and replacing it with $A_0^\d = \d_0(A_0, H_0, \epsilon_0)$.
This assignment generates counterfactual data $H_1(A_0^\d)$
and $A_1(A_0^\d)$, where we use notation $X(A_0^\d)$ to denote the counterfactual value of $X$ that would have been observed had treatment at time 0, $A_0$ been assigned according to $\d$. $H_1(A_0^\d)$ is referred to as the
counterfactual history and $A_1(A_0^\d)$ is referred to as the
\textit{natural value of treatment} \citep{robins2004effects,
  richardson2013single, YoungHernanRobins2014}, i.e., the value that
treatment would have taken if the intervention is performed at time
$t=0$ but discontinued thereafter.
At time $t=1$, the intervention is
likewise defined by a function $\d_1(a_1, h_1, \epsilon_1)$. However,
at $t=1$ (and all subsequent times if there are more than two time
points), the function must be applied 
applied to both the
natural value of treatment \emph{and} the counterfactual history. That is, at time $t=1$, the intervention
is
defined by removing node $A_1$ from the DAG and replacing it with
$A_1^\d = \d_1(A_1(A_0^\d), H_1(A_0^\d), \epsilon_1)$.

We refer to these longitudinal interventions, and the subsequent methods to identify and estimate effects under such interventions, as LMTPs. 
We now give examples of how the functions $\d_t$ may
be defined, explain how they mathematically generalize static and dynamic 
interventions, and discuss novel and useful interventions
that may be defined using this setup. 


\section{A review of static and dynamic interventions}

The function $\d_t(a_t, h_t, \epsilon_t)$ that defines the
intervention or treatment policy can be categorized by the inputs it non-trivially depends on. An intervention which depends on no inputs, i.e. is applied as a constant across all study units, is a static intervention. For this tutorial, we refer to an intervention which depends only on a study unit's past covariates as a dynamic intervention. Finally, we call an intervention which depends on a study unit's natural value of treatment (and possibly past covariates, and possibly a randomizer $\epsilon_t$) an MTP. We summarize these hierarchical categories
in Table \ref{tab:interventions}. 

\subsection{Static interventions}

In a static intervention, $\d_t(a_t, h_t, \epsilon_t)$ is a
constant value, i.e., it does not actually vary with $a_t$, $h_t$, nor
$\epsilon_t$.
\begin{example}[Average treatment effect]\label{exstatic}
  For the two time point example, one might examine the counterfactual
  outcomes in a hypothetical world in which all units are treated at
  both time points ($\d_t=1$ for $t\in\{0,1\}$), and contrast them to
  a hypothetical world in which no units are treated at either time
  point ($\d_t=0$ for $t\in\{0,1\}$), giving rise to the well-known
  average treatment effect.
\end{example}

\subsection{Dynamic interventions}
In a dynamic intervention, the function $\d_t$ assigns a treatment
value according to a unit's covariate history $h_t$, but
does not vary with $a_t$ nor
$\epsilon_t$. 
This is often used in observational studies when study units need to meet an indication of interest for a treatment or policy to reasonably begin, for example, a severity of illness indicator or socioeconomic threshold. 


\begin{example}[Corticosteroids for COVID-19 hospitalized patients]\label{eq:excort}
  
  One dynamic treatment regime application
  is to study the effect of initiating a corticosteroids
  regimen for COVID-19 patients \citep{hoffman2022comparison}. For example, we might estimate mortality under a hypothetical policy where corticosteroids are administered for six days if and when a COVID-19 patient first meets a severity of illness criteria (i.e. low levels of blood oxygen). In notation,
  
  

  \[\d_t(h_t)=\begin{cases}
      1 &\text{ if } l_s^*=1 \text{ for any } s\in\{t-5,\ldots, t\}\\
      0&\text{ otherwise,}
    \end{cases}\]
      where $L_t^*$ is a variable in $H_s$ that denotes the first instance of low levels of blood oxygen.
 \end{example}

\section{Modified treatment policies}

In contrast to static and dynamic interventions, a modified treatment policy, in which the intervention function
$\d_t(a_t, h_t, \epsilon_t)$ non-trivially depends on the natural value of treatment $a_t$, and perhaps on $h_t$ and/or $\epsilon_t$.

\begin{example}[Threshold intervention]\label{exthreshold}

One example of an MTP is a threshold function, where all natural exposure values which fall outside of a certain boundary are intervened upon to meet a constant value. This type of intervention could be used to assess the effect of lifestyle interventions, for example, intervening on individuals' average number of drinks per week and estimating
the risk of coronary heart disease \citep{taubman2009intervening}. If we categorize drinks per week as 1 = "none," 2 = "1-5," 3 = "6-10,"  4 = "11-15," and  5 = ">25", and we intervene to lower all individuals in the highest two drinks-per-week categories to "6-10," we can consider that intervention in notation as,

\[\d_t(a_t)=\begin{cases}
      a_t & \text{ if } a_t < 4\\
      3 & \text{ otherwise. }
    \end{cases}\]
    
\end{example}

\begin{example}[Smoking cessation policy]\label{exsmoking}

  Another example is a hypothetical policy in
  which half of all current smokers quit smoking forever \citep{robins2004effects}. This intervention is motivated by the infeasibility of studying a world in which all current smokers quit smoking forever, since genetics, environment, and many other factors (likely unmeasured) will always create some portion of current smokers who will never quit. Letting
  $A_t$ denote a random variable denoting smoking and $\epsilon_t$ a random draw from a uniform
  distribution in $(0,1)$, this intervention
  may be represented in notation as
  \[\d_t(a_t,\epsilon_t)=\begin{cases}
      0 & \text{ if } \epsilon_t<0.5 \text{ and } a_t=1\\
      a_t & \text{ otherwise, }
    \end{cases}\].  
\end{example}


    


\begin{example}[Multiplicative or additive shift functions]\label{exshift}

  Shift functions assign treatment by modifying the natural value of
  the exposure by some constant $\delta$. This intervention can be
additive onto the exposure value, such as estimating the effect of a hypothetical intervention to reduce lung cancer resection surgeries lasting longer than 60 minutes by 15 minutes \citep{haneuse2013estimation}.

    \[\d_t(a_t)=\begin{cases}
      a_t & \text{ if } a_t \leq 60 \\
      a_t - 15 & \text{ otherwise. }
    \end{cases}\] 
\noindent This shift function could also change the exposure on a
  multiplicative scale.  For example, we
  may be interested in studying the effect of an intervention which
  doubles the number of street lights for roads with less than 10
  lights per mile on nighttime automobile accidents.

  \[\d_t(a_t)=\begin{cases}
      a_t & \text{ if } a_t \ge 10 \\
      2a_t & \text{ otherwise. }
    \end{cases}\]

\end{example}

\noindent We provide additional examples of interesting 
interventions in the Appendix. Readers should also note that some interventions we refer to as MTPs \citep{haneuse2013estimation} in this tutorial are called interventions which depend on the natural value of treatment, \citep{YoungHernanRobins2014} and are related to methods for stochastic treatment regimes which shift the natural population- or individual-level values of treatment \citep{munoz2012population, van2018targeted} discussed in other literature.

\section{Causal estimands and identifying parameter}

Once an intervention is specified, the counterfactual outcomes of
observations under a specific $\d$ are denoted as
$Y(\bar{A}_\tau^\d)$, where $\bar{A}$ indicates the history of measurements of $A$ for all time points, i.e. $\bar A = (A_1, \ldots, A_\tau)$. Causal effects are defined as a distribution of contrasts of $Y(\bar{A}_\tau^\d)$ under different interventions, $\d'$ and
$\d^\star$. In this tutorial, we focus on
$\E[Y(\bar{A}_\tau^{\d'}) - Y(\bar{A}_\tau^{\d ^ {\star}})]$ as our causal estimand of interest. The
functions $\d'$ and $\d^\star$ may be any type of intervention, including ``no intervention.'' 


The next step in a formal causal inference analysis is to write the
counterfactual expectation $\E[Y(\bar{A}_\tau^{\d'})]$ as a formula that depends only
on the observed data distribution---i.e., an
identifying formula. This will generally require assumptions, some of
which are untestable with the data available. The mathematically rigorous form of the
assumptions is given elsewhere \citep{richardson2013single,
  diaz_nonparametric_2021}, but we state them here in simple terms:

\begin{assumption}[Positivity or common support \citep{YoungHernanRobins2014}]
  If it is possible to find an observation with history $h_t$ with an
  exposure of $a_t$, then it is also possible to find an observation
  with history $h_t$ with an exposure $\d(a_t, h_t, \epsilon_t)$.
\end{assumption}
\begin{assumption}[Strong sequential randomization\citep{diaz_nonparametric_2021}]
  This assumption states
  that all common causes of the intervention variable $A_t$ and $(U_{L,t+1}, U_{A,t+1})$ are measured and recorded in $H_t$. 
  This is
  generally satisfied if $H_t$ contains all common causes of $A_t$ and
  $(L_{t+1}, A_{t+1}, \ldots, L_\tau, A_\tau,\allowbreak Y)$, where $\tau$ is
  the last time point in the study.
\end{assumption}

\begin{assumption}[Weak sequential randomization\citep{richardson2013single, diaz_nonparametric_2021}] 
  This assumption states
 that all common causes of the intervention variable $A_t$ and $(U_{L,t+1})$ are measured and recorded in $H_t$ \citep{taubman2009intervening}, 
  and is
  generally satisfied if $H_t$ contains all common causes of $A_t$ and
  $(L_{t+1}, \ldots, L_\tau, \allowbreak Y)$.
\end{assumption}


Identification of LMTPs requires the strong version of sequential randomization. Interventions that do not depend on the natural value of treatment, such as static and dynamic interventions, require the weak version of sequential randomization.

\subsection{Positivity}


Violations to positivity can be \textit{structural}, meaning there are certain characteristics of an individual or unit which will never yield receipt of the treatment assignment under the intervention. This type of positivity violation will not improve even with an infinite sample size. Violations to positivity can also be \textit{practical}, meaning due to random chance or small datasets, there are certain covariate combinations with zero or near-zero predicted probabilities of treatment. For time-varying treatments, this positive probability must be maintained at each time point \citep{petersen2012diagnosing}. Practical positivity violations can increase the finite bias and variance of estimates and severely threaten the validity of casual inference analyses when not addressed \citep{petersen2012diagnosing}. By design, non-static interventions (e.g. dynamic treatment rules, MTPs) may help define estimands with plausible positivity, since the function $\d$ can be modified to affect the exposure of only observations which are not subject to positivity violations, either structurally or practically.


This can be seen in the interventions described above, for instance, the additive shift in Example \ref{exshift}. We can conceptualize a world in which a continuous exposure is instead observed at some fixed value higher or lower than it was factually observed for every unit in the study; for example, if surgery times were 15 minutes shorter for all lung resection biopsies. However, this type of uniform hypothetical modification is destined for structural positivity violations, because at the lowest end of the observed exposure range, there will by definition be no support for the intervened exposure level $\d(a_t)$ (much less conditional on the observation's history $h_t$).
This can be avoided by constraining the range of $a_t$ affected by the hypothetical intervention, so that no $\d(a_t)$ values are produced outside the
observed range of
$A$. The intervention function can also be modified to accommodate any other remaining structural or practical positivity violations. For example, clinical knowledge may inform us that a treatment of interest will never be administered after a certain amount of time since a disease diagnosis has passed, so the hypothetical intervention would restrict the values of $t$ in which the intervention can occur. Alternatively, if there is not enough support in the data for individuals of a certain covariate strata or at a naturally observed exposure level to receive the intervention, a different estimand can be defined.

\subsection{Identification formula}

Under Assumptions 1 and 2, or 1 and 3, the estimand is identified by the
generalized g-formula \citep{Robins86}. A re-expression of this generalized g-formula \citep{bang2005doubly, diaz_nonparametric_2021}  involves recursively defining the expected outcome under the
intervention, conditional on the observation's observed exposure and
history, beginning at the final time point, and proceeding until the
earliest time point. We illustrate the
g-formula for two time points below:

\begin{enumerate}

\item Start with the conditional expectation of the outcome $Y$ given $A_1=a_1$ and $H_1=h_1$. Let this function be denoted $Q_1(a_1, h_1)$.
\item Evaluate the above conditional 
  expectation of $Y$ if $A_1$ were changed to $A^\d_1$, which results in 
  a pseudo outcome $\tilde Y_1=Q_1(A^\d_1, H_1)$.
\item Let the true expectation of $\tilde Y_1$ conditional on
  $A_0=a_0$ and $H_0=h_0$ be denoted $Q_0(a_0, h_0)$.
\item Evaluate the above
  expectation of $\tilde Y_1$ if $A_0$ were changed to $A^\d_0$, which results in
  $\tilde Y_0=Q_0(A^\d_0, H_0)$.
\item Under the identifying assumptions, we have
  $\E[Y(\bar{A}_\tau^\d)]=\E[\tilde Y_0]$.
  
\end{enumerate}

\section{Estimation}

Once a causal estimand is defined and identified, the researcher's task is to estimate the statistical quantity, e.g. $\E[\Tilde{Y}_0]$. We
now discuss several estimators, both parametric and non-parametric, and provide their algorithm steps in pseudo-R code in the Appendix.

\subsection{Parametric estimation}

 The simplest option for estimation is to fit parametric outcome regressions for each step of the g-formula identification result. This ``plug-in'' esimator is often referred to as the parametric g-formula or g-computation. 
 Another option is to use an estimator which relies on the exposure mechanism, for example, the Inverse Probability Weighting (IPW) estimator. IPW estimation involves reweighting the observed outcome by a quantity which represents the likelihood the intervention was received, conditional on covariates. 

Obtaining point estimations with the g-computation and IPW algorithms is computationally straightforward. If the exposure regression for IPW or outcome regression for
g-computation are estimated using pre-specified parametric statistical models, standard errors for the estimate can be computed using bootstrapping or the Delta method. However, in causal models with large numbers of covariates and/or complex mathematical relations between confounders, exposures, and
outcomes, parametric models are hard to pre-specify, and they are unlikely to consistently estimate the regressions. If the regression for the outcome (for g-computation) or treatment (for IPW) are misspecified, the final estimates will be biased.

One way to alleviate model misspecification is to use flexible approaches which incorporate model selection (e.g. 
LASSO, splines, boosting, random forests, ensembles thereof, etc.) to estimate the exposure or outcome
regressions. Unfortunately, there is generally not statistical theory to support the standard errors of the g-computation or IPW estimators with such data-adaptive regressions. Standard inferential tools such as the bootstrap will fail because these estimators generally do not have an asymptotically normal distribution after using data-adaptive regressions \citep{van2011targeted}. Thus, other methods are needed to accommodate both model selection and flexible regression techniques while still allowing for statistical inference.

\subsection{Non-parametric estimation}

Here we will advocate for the use of general, non-parametric estimators for LMTPs \citep{diaz_nonparametric_2021}.
 These estimators use both an exposure and outcome regression, and allow the use of machine learning to estimate the regressions while still obtaining valid statistical uncertainty quantification on the final estimates. 
  These estimators also remain consistent under inconsistent estimation of at most one of the nuisance parameters. 

 The two non-parametric estimators 
 encoded in the R package \emph{lmtp} \citep{lmtppkg, nickpaper} are Targeted Maximum Likelihood Estimation (TMLE) \citep{van2011targeted, van2012targeted, diaz_nonparametric_2021} and Sequentially Doubly Robust (SDR) estimation \citep{luedtke2017sequential, rotnitzky2017multiply, diaz_nonparametric_2021}. A third non-parametric estimator, iterative TMLE (iTMLE), is not encoded in the R package but could be adapted from \citet{luedtke2017sequential}.  TMLE is a doubly robust estimator for a time-varying treatment in the sense that it is consistent as long as all outcome regressions for times $t>s$ are consistently estimated, and all treatment mechanisms for times $t\leq s$ are consistent, for some time $s$. In contrast, SDR and iTMLE are sequentially doubly robust in that they are consistent if for all times $t$, either the outcome or the treatment mechanism are consistently estimated \citep{luedtke2017sequential, diaz_nonparametric_2021}. Since TMLE and iTMLE are substitution estimators, they are guaranteed to produce estimates which remain within the observed outcome range. SDR and iTMLE produce estimators with more robustness than TMLE.
 
 Table \ref{tab:est-props} compares the statistical properties of various
estimators. We provide practical guidance for choosing between estimation techniques in the Appendix.
Of note, for the statistical properties of these estimators to hold, certain technical requirements must be met. These requirements are 
detailed in the Appendix. All of the aforementioned examples meet these requirements.


\section{LMTP applications in practice}

In the worked example, we estimate the effect of delaying intubation
on mortality. Other examples of LMTP applications in similar
populations include studying the effect of a delay in intubation on an
outcome of acute kidney injury, where death is a competing risk
\citep{diaz2022causal}, or studying an intervention on a continuous
measure of hypoxia in acute respiratory distress patients
\citep{diaz_nonparametric_2021}. LMTP also accommodates interventions
involving multiple treatments, such as delaying intubation by one day
\emph{and} increasing fluid
intake. 

There are multiple other examples of researchers using LMTPs to answer real-world problems. \citet{nugent2021evaluating} used an LMTP to study the effects of mobility on COVID-19 case rates. Specifically, they studied the effect of a longitudinal modified shift on the observed mobility distribution on the number of newly reported cases per 100,000 residents two weeks ahead. A similar analysis could be done to look at interventions on masking or vaccination policies. Mobility, masking, and vaccination are important examples of when static or dynamic policy estimands may be unappealing because of the geographical, political, and cultural variation that exists even within relatively small regions. 
This applies to environmental exposures and health policies as well. For instance, \citet{rudolph2022effects} 
used LMTPs to estimate the effects of Naxolone access laws on opioid overdose rates. 

Other examples of LMTP applications include \citet{jafarzadeh2022relationship}'s study of the effect of an intervention on knee pain scores over time on an outcome of knee replacement surgery. \citet{huling2022public} investigated the effects of interventions to public health nursing on the behaviors of clients in the Colorado Nurse Support Program. \citet{mehta2021primary} researched whether an increase in primary care physicians has an effect on post-operative outcomes in patients undergoing elective total joint replacement. 
Tables \ref{tab:bin-table} and \ref{tab:cont-table} show additional applications of LMTP applications, as well as examples of possible static and dynamic treatment rules answering similar research questions.

\section{Illustrative example}

\subsection{Motivation}

In the following application, we study a clinical question, ``what is the effect of delaying intubation of invasive mechanical ventilation (IMV) on mortality for hospitalized COVID-19 patients in New York City's first COVID-19 wave?" 
Studying the effect of intubation timing is particularly ill-suited for static interventions because there is no scenario in which intubation at a certain study time could be uniformly applied across all critically ill patients. 
A dynamic intervention, which could help to evaluate a world in which patients are intubated when they meet a certain severity threshold, e.g. an oxygen saturation breakpoint, is less clinically relevant because intubation guidance may vary considerably between providers dependent on training, hospital policies, and ventilator availability \citep{tobin2020caution, perkins2020recovery}. 
For this reason, an LMTP which varies the natural time of intubation by a minimal amount (for example, one day), may be a more realistic hypothetical intervention to study when considering the mechanistic effect of a delay-in-intubation strategy on mortality.
 
\subsection{Measures and analysis}

The population of interest is adult patients hospitalized and diagnosed with COVID-19 during Spring 2020.  The cohort contains 3,059 patients who were admitted to NewYork-Presbyterian's Cornell, Queens, and Lower Manhattan locations between March 1-May 15, 2020 \citep{goyal2020clinical, schenck2021critical, diaz2022causal}. This research was approved with a waiver of informed consent by the Weill Cornell Medicine Institutional Review Board (IRB) 20-04021909.

The exposure of interest is the maximum level of daily supplemental oxygen support. This can take three categories, 0: no supplemental oxygen, 1: non-IMV supplemental oxygen support, and 2: IMV. The intervention of interest describes a hypothetical world in which patients who require IMV have their intubation delayed by one day, and instead receive non-IMV supplemental oxygen support on the observed day of intubation.

\begin{equation*}\label{eq:exmtp}
\d_t(a_t,h_t) =
  \begin{cases}
    1 &\text{ if } a_t=2 \text{ and } a_s \leq 1 \text{ for all } s < t,\\
   a_t & \text{ otherwise}
 \end{cases}
\end{equation*}

\noindent Since patients in this observational data are subject to loss to follow up via discharge or transfer, the intervention additionally includes observing patients through all 14 days from hospitalization. 
    
The primary outcome is time to death within 14 days from hospitalization. 
 Hospital discharge or transfer to an external hospital system is considered an informative loss to follow-up. The causal estimand of interest is the difference in 14-day mortality rates between a hypothetical world in which there was a one-day delay in intubation and no loss to follow-up, and a hypothetical world in which there was no loss to follow-up and no delay in intubation.

Since the intervention is an LMTP, we require positivity and strong sequential randomization to identify our parameter of interest. The common causes we assume to satisfy the latter requirement include 37 baseline confounders and
14 time-varying confounders per time point.

The R package \emph{lmtp} \citep{lmtppkg, nickpaper} was used to
obtain estimates of the difference in 14-day mortality rates under the
two proposed interventions using SDR estimation. A 
superlearner \citep{Wolpert1992, Breiman1996,
  vanderLaanPolleyHubbard07} library of various candidate learners was
used to estimate the intervention and outcome estimation. We demonstrate code on synthetic data 
at 
  \href{https://github.com/kathoffman/lmtp-tutorial}{https://github.com/kathoffman/lmtp-tutorial}
  and provide several additional details in the Appendix. 

\subsection{Results}

\begin{figure}[H]
\caption{Panel A: Estimated incidence of mortality between a Delayed Intubation MTP (blue) and No intervention (red). Panel B: Estimated incidence difference in mortality if the Delayed Intubation MTP were implemented during Spring 2020. In both panels, 95\% simultaneous confidence bands \citep{westling2020correcting} cover the sets of point estimates.}
\centering
\includegraphics[width=17cm]{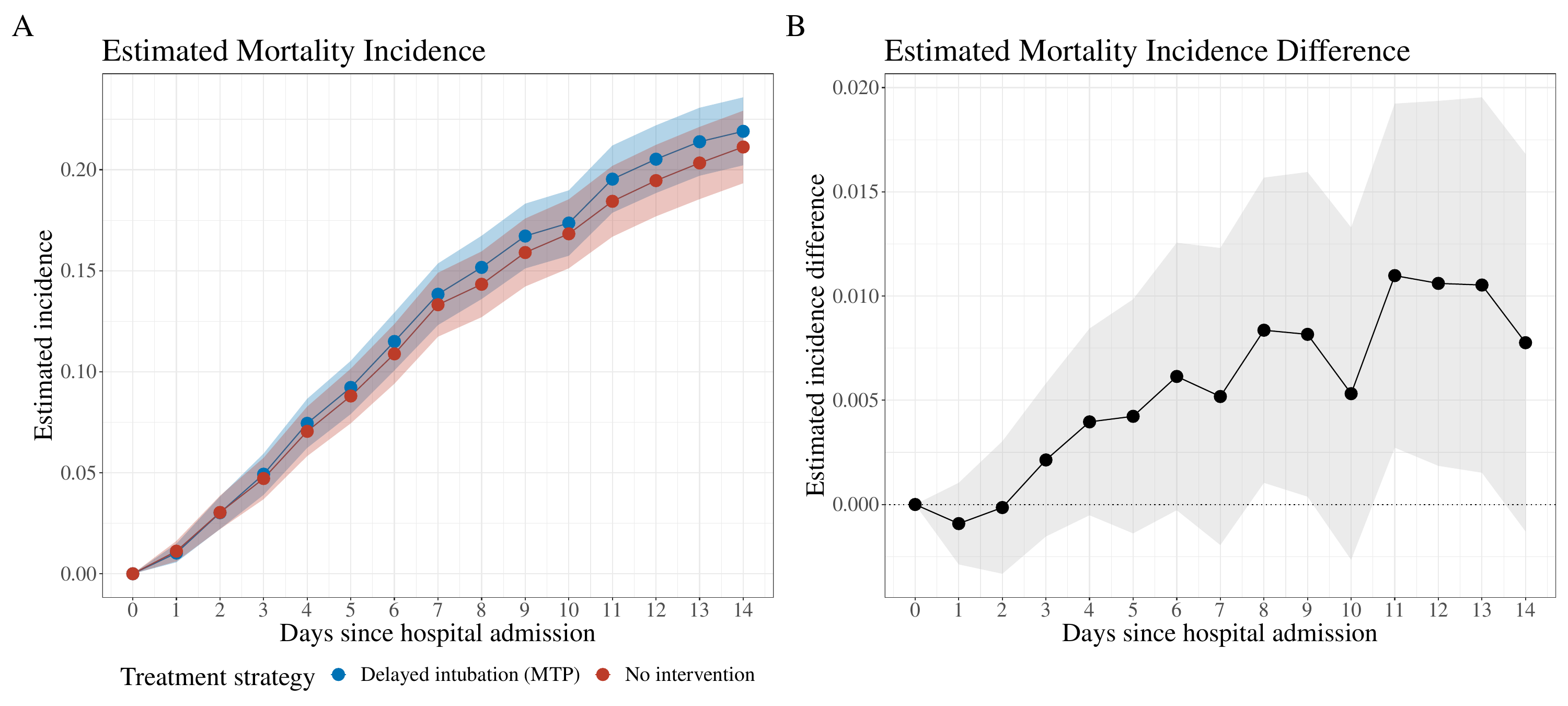}\label{fig1}
\end{figure}

The estimated 14-day mortality incidence under no intervention on intubation was 0.211 (95\% CI 0.193-0.229). The same incidence under a hypothetical LMTP in which intubation were delayed by 1 day was
0.219 (0.202-0.236). The estimated incidences across all time points are shown in Figure \ref{fig1}. 

\section{Discussion}

The LMTP methodology allows researchers to define,
identify, and estimate scientifically relevant estimands, including those that involve challenges such as loss-to-follow-up, survival analysis,
missing exposures, competing risks, and 
interventions that include multiple exposures and/or continuous exposures. LMTPs are particularly useful for designing estimands with a higher chance of meeting the positivity assumption. In
addition, the existing packages that implement doubly and sequentially robust estimators
enable researchers to take advantage of statistical learning
algorithms to estimate the intervention and outcome mechanisms, thereby increasing the likelihood of estimator consistency.

 
While the LMTP methodology expands the applied researcher's toolbox to accommodate longitudinal interventions which non-trivially depend on the natural value of treatment, there are considerations and
limitations to its implementation in real-world applications. First,
LMTPs (as well as static and dynamic interventions in a longitudinal setting) require discretizing time
over intervals. Depending on the data collection process, this may
cause issues in temporality or loss of data granularity. This discretization may also cause issues in small sample sizes if there are very few outcomes within a certain time point and the applied researcher hopes to estimate the outcome regressions using any statistical learning algorithm which segments the data for training/testing. Second,
although LMTPs may be used to formulate alternative estimands that satisfy the respective positivity
assumption, the alternative estimand must be 
scientifically relevant. 
In addition, practical positivity violations may still occur, and possible solutions, such as
truncating the density ratios at a certain threshold, are arbitrary and the potential for bias is unclear. Third, some particular estimator applications may be computationally intensive.  
Despite these limitations, we hope
the explanation of LMTPs, illustrative example, and corresponding Github
repository are a useful toolset for researchers hoping to implement
LMTPs in their applied work.


 
 \section{Tables}
 
\begin{table}[H]
\begin{threeparttable}
\caption{An overview of common intervention types in the causal inference literature and whether they are estimable using the \emph{lmtp} R package.}
\centering
\begin{tabular}{  m{11cm} m{4.5cm} } 
\toprule
  \textbf{Intervention and definition} & 
\textbf{Estimable with \emph{lmtp}?} \\
  \midrule
  \textbf{Static:} all units receive the same treatment assignment  & Yes\tnote{a} \\ 
  \midrule
  \textbf{Dynamic:} A unit’s treatment assignment is determined according to their covariate history\tnote{b} 
 & Yes\tnote{a}  \\ 
   \midrule
  \textbf{Modified:} A unit’s treatment assignment is determined according to their natural exposure value, and possibly their covariate history\tnote{b,c} & See assumptions\tnote{d}  \\ 
  \hline
\end{tabular}
\begin{tablenotes}
 \item [a] The software should only be used with discrete exposures. The software will output a result with a continuous exposure, but this estimator will not have good statistical properties.
 \item [b] The covariate history could include prior exposure values under the intervention function.
\item [c] An MTP may also depend on a randomizer.
     \item [d]Assumptions require the intervention function $\d$ does not depend on the distribution $\P$, and,
one of: (1) the exposure is discrete or (2) the exposure is continuous but $\d$ satisfies piecewise smooth invertibility. See Technical Requirements in Appendix for more details.
\end{tablenotes}
\label{tab:interventions}
\end{threeparttable}
\end{table}

\begin{table}[H]
\begin{threeparttable}
\caption{Comparison of statistical properties for five possible estimators for LMTPs: g-computation (G-COMP), inverse probability weighting (IPW), targeted maximum likelihood estimation (TMLE), sequentially doubly robust estimation (SDR), and iterative TMLE (iTMLE).}
\centering

\begin{tabular}[t]{l c c c c c}
\toprule
\textbf{Statistical Property} & \textbf{G-COMP} & \textbf{IPW} & \textbf{TMLE} & \textbf{SDR} & \textbf{iTMLE}\tnote{a} \\
\midrule
Uses outcome regression & \usym{1F5F8} &  & \usym{1F5F8} & \usym{1F5F8} & \usym{1F5F8}\\
\midrule
Uses treatment regression &  & \usym{1F5F8} & \usym{1F5F8} & \usym{1F5F8} & \usym{1F5F8} \\
\midrule
Doubly robust\tnote{b}  &  &  & \usym{1F5F8} & \usym{1F5F8} & \usym{1F5F8}\\
\midrule
Sequentially doubly robust\tnote{c}  &  &  &  & \usym{1F5F8} & \usym{1F5F8}\\
\midrule
\makecell{Valid inference\tnote{d}\\ using parametric regressions\\ (i.e. generalized linear models)} & \usym{1F5F8} & \usym{1F5F8} & \usym{1F5F8} & \usym{1F5F8} & \usym{1F5F8} \\
\midrule
\makecell{Valid inference\tnote{d}\\ using data-adaptive regressions\\ (i.e. machine learning)} &  &  & \usym{1F5F8} & \usym{1F5F8} & \usym{1F5F8} \\
\midrule
\makecell{Guaranteed to stay within\\ observed outcome range} & \usym{1F5F8} &  & \usym{1F5F8} &   & \usym{1F5F8} \\
\bottomrule
\end{tabular}
\begin{tablenotes}
     \item [a] The iTMLE estimator is not currently available within the R package lmtp.
     \item [b] The estimator is consistent as long as all outcome regressions for times $t>s$ are consistently estimated, and all treatment mechanisms for times $t\leq s$ are consistent, for some time $s$.
     
     \item [c] The estimator is consistent if, for every time point, either the outcome regression or the treatment mechanism is consistently estimated.
     
     \item [d] Includes standard errors, confidence intervals, and p-values.
\end{tablenotes}
\label{tab:est-props}
\end{threeparttable}
\end{table}

\newpage

\begin{table}[H]
\begin{threeparttable}
\small
\caption{Examples of static, dynamic, 
and modified interventions for (1) binary and (2) continuous point-in-time exposures. These can be expanded to any categorical exposure. We denote a random variable drawn from Bernoulli distribution with probability $0.5$ with $\epsilon$ unless otherwise noted.}
\centering

\begin{tabular}[t]{p{2.5cm}p{2cm}p{4cm}p{6cm}}
\toprule \thead{Exposure} & \thead{Intervention} & \thead{``What if...''} & 
 \thead{Shift Notation}\\ 
\midrule

\multirow{2}{*}{\makecell{Point-in-time \\ Binary, \\ e.g. vaping \\ $(a=1)$}} & Static & \makecell{no one vapes} & 
    $\d(a, h, \epsilon) = 0$ \\

  \cmidrule{2-4}
   
    & Dynamic  & \makecell{only those working \\ non-standard work \\ hours $(l^*=1)$ vape} &  \makecell{$\d(a, h, \epsilon)=\begin{cases} 1  &\text{ if } l^*=1 \\ 0  &\text{ otherwise} \end{cases}$} \\

     \cmidrule{2-4}



    & Modified & \makecell{a random half of \\ current vapers stop \\ vaping} & \makecell{$\d(a, h, \epsilon) = \begin{cases} \epsilon &\text{ if } a=1\\
        a &\text{ otherwise } \end{cases}$} \\
  
\midrule

\multirow{1}{*}{\parbox{3cm}{\raggedright Point-in-time Continuous, \\ e.g. exposure \\ to pollution as \\ measured by \\ the Air Quality \\ Index (AQI)  \\scale}} & Static & \makecell{all counties are \\ exposed to an AQI \\ of 10} & \makecell{$\d(a, h, \epsilon) = 10$} \\

   \cmidrule{2-4}

    & Dynamic  & \makecell{all urban ($l^*=1$) \\ counties are exposed \\ to an AQI of 40 and  \\ all rural ($l^*=0$) \\ counties are exposed to \\ an AQI of 20} & \makecell{$\d(a, h, \epsilon) = \begin{cases}  40 &\text{ if } l^*=1 \\
        20 &\text{ otherwise} \end{cases}$} \\

         \cmidrule{2-4}



    & Modified & \makecell{all counties with an \\ AQI higher than 20 are \\ exposed to an AQI \\ 10\% lower than what \\ they were naturally \\ exposed to} & \makecell{$\d(a, h, \epsilon) = \begin{cases} a \times 0.9 &\text{ if } a > 20 \\ a &\text{ otherwise} \end{cases}$} \\

\bottomrule
\end{tabular}
\begin{tablenotes}
\end{tablenotes}
\label{tab:bin-table}
\end{threeparttable}
\end{table}

\newpage

\begin{table}[H]
\begin{threeparttable}
\small
\caption{Examples of static, dynamic, and modified interventions for a binary time-varying exposure. We index study time at $t=0$ and denote a random variable drawn from Bernoulli distribution with probability $0.5$ with $\epsilon_t$.}
\centering

\begin{tabular}[t]{p{2.5cm}p{2cm}p{3cm}p{8.5cm}}
\toprule \thead{Exposure} & \thead{Intervention} & \thead{``What if...''} &
 \thead{Shift Notation}\\ 
\midrule

\multirow{2}{*}{\makecell{Time-varying \\ Binary, e.g. \\ corticosteroids \\ 
 receipt $(a_t=1)$}} & Static & \makecell{all patients receive \\ corticosteroids for \\ the first 6 days of \\ hospitalization} & \makecell{$\d_t(a_t,h_t,\epsilon_t) = \begin{cases} 1 &\text{ if } t \leq 5 \\ 
    0 &\text{ otherwise} \end{cases} $} \\

  \cmidrule{2-4}

   & Dynamic  & \makecell{patients receive \\ corticosteroids \\ for 6 days once \\ they become \\ hypoxic $(l^*_t=1)$} &
      \makecell{$\d_t(a_t,h_t,\epsilon_t) = \begin{cases} 1 &\text{ if } l^*_s=1 \text{ for } s\in\{t-5,\ldots, t\} \\ 
    0 &\text{ otherwise} \end{cases} $} \\

        \cmidrule{2-4}

  

  & Modified &
 \makecell{patients' receipt of \\ corticosteroids is \\ delayed by 1 day} & 
 \makecell{$\d_t(a_t,h_t,\epsilon_t) = \begin{cases} 0 &\text{ if } a_t=1 \text{ and } a_{t-1}=0\\ 
    a_t &\text{ otherwise} \end{cases} $} \\

\bottomrule
\end{tabular}
\begin{tablenotes}
\end{tablenotes}
\label{tab:cont-table}
\end{threeparttable}
\end{table}

\newpage


\bibliography{refs}

\end{document}


\section{Appendix}\label{appendix}

\subsection{Additional Intervention Examples}

\subsubsection{Stochastic Interventions}

\begin{example}[Incremental propensity score interventions based on
  the odds ratio]\label{exkennedy}
  A recently developed type of stochastic interventions is
  \citet{kennedy2019nonparametric}'s incremental propensity score
  interventions (IPSI), which shifts a unit's probability of receiving
  treatment conditional on their according to some constant
  $\delta$. First, a shifted propensity score is defined:
\begin{equation*}
  \pi^\d_t(h_t) = \frac{\delta\pi_t(h_t)}{\delta\pi_t(h_t) + 1 - \pi_t(h_t)}
\end{equation*}
Then, a random variable $\epsilon_t$ is drawn from a uniform
distribution in $(0,1)$ and the intervention is defined as:
\begin{equation*}
\d_t(h_t, \epsilon_t, \pi_t) = I(\epsilon_t < \pi^\d_t(h_t)),
\end{equation*}
This IPSI is said to be based on an odds ratio because the odds ratio
of $\pi^\d_t(h_t)$ to $\pi_t(h_t)$ is equal to
$\delta$. \citet{naimi2021incremental} used an IPSI to study the
causal relationship between vegetable density consumption and the risk
of preeclampsia among pregnant women. They studied whether
preeclampsia would increase if women's propensity of eating a minimum
amount of vegetables increased by an odds ratio of $\delta$, for
example, $\delta = 1.5$ would mean a woman with a propensity of 0.35
increases to 0.45.
\end{example}



\subsubsection{Modified Treatment Policies}

\begin{example}[Incremental propensity score based on the risk
  ratio]\label{exwen}
  Another type of IPSI proposed by \citet{wen2021intervention} begins
  with a slightly different set-up than \citet{kennedy2019nonparametric}. Instead of relying on the
  treatment mechanism, there is a random draw $\epsilon_t$ from a
  Uniform $(0,1)$ distribution. If the draw is less than some
  $\delta$, then the treatment assignment is the natural value of
  treatment. If not, it is some constant value, for example 0.
  \begin{equation*}
    \text{$\d_t(a_t,\epsilon_t)$} = \begin{cases}
      \text{$a_t$ if $\epsilon_t < \delta$},\\
      \text{$0$ otherwise}
    \end{cases}
  \end{equation*}
  This intervention is said to be based on the risk ratio because $\delta = \pi^\d_t(h_t)/\pi_t(h_t)$, where $\pi^\d_t(h_t)$ is the post-intervention propensity score and $\pi_t(h_t)$ is the propensity score in the observed data generating mechanism. \citet{wen2021intervention} argue the risk ratio interpretation may be more intuitive for collaborators, and they demonstrate this IPSI in an application studying the effect of PrEP usage increases on sexually transmitted infection rates.

  \end{example}

\subsection{Estimation Algorithms}

\subsubsection{G-Computation Algorithm}

The G-Computation algorithm steps and pseudo-code for a simple example with two time points is shown below. 

\begin{enumerate}
\item Fit a generalized linear model (GLM) for $Y$ conditional on $A_1=a_1$ and
  $H_1=h_1$. Call this $\hat{Q_1}(a_1, h_1)$.
  
  \texttt{Q1\_hat} $\gets$ \texttt{glm}$(\text{outcome} = Y, \text{predictors} = \{H_1, A_1\})$

\item Modify the data set used in step (1) so that the values in the column for $A_1$ are changed to $A^\d_1$. Obtain the predictions for the model $\hat{Q_1}$ using this modified data set. These are pseudo-outcomes
  $\tilde Y_1$.

\texttt{pseudo\_Y1} $\gets$ \texttt{predict}$(\text{fit} =   \texttt{Q1\_hat}, \text{new data} = \{H_1, A_1 = A_1^\d\} )$
  
\item Fit a generalized linear model (GLM) for $\tilde Y_1$ conditional on
  $A_0=a_0$ and $H_0=h_0$. Call this $\hat{Q_0}(a_0, h_0)$.

  \texttt{Q0\_hat} $\gets$ \texttt{glm}$(\text{outcome} = \texttt{pseudo\_Y1}, \text{predictors} = \{H_0, A_0\})$
  
\item  Modify the data set used in step (3) so that the values in the column for $A_0$ are changed to $A^\d_0$. Obtain the predictions for the model $\hat{Q_0}$ using this modified data set. These are pseudo-outcomes
  $\tilde Y_0$.

\texttt{pseudo\_Y0} $\gets$ \texttt{predict}$(\text{fit} =   \texttt{Q1\_hat}, \text{new data} = \{H_0, A_0 = A_0^\d\} )$
  
\item Average $\tilde Y_0$, i.e. compute $\hat\E[\tilde Y_0]$.

\texttt{estimate} $\gets$ \texttt{mean(pseudo\_Y0)}
\end{enumerate}

\subsubsection{Inverse Probability Weighting Algorithm}

The IPW estimator relies on a density ratio $\r_t = \g_t^\d(a_t\mid h_t)/\g_t(a_t\mid h_t)$, where $\g_t^\d(a_t\mid h_t)$ is the density of the intervened exposure, and $\g_t(a_t\mid h_t)$ is the density of the naturally observed exposure. Practically, $\r_t$ can be computed using a clever classification trick proposed in \citet{qin1998inferences, cheng2004semiparametric} and utilized in \citet{diaz_nonparametric_2021}. The IPW algorithm with pseudo-code for a simple example with two time points is as follows. 

\begin{enumerate}
\item Duplicate each row of the data set so that there are $2n$ rows. The first row of a duplicated pair should contain observed values $A_1$, and the second row should be modified so that $A_1 = A_1^\d$. A new column $\Lambda_1$ should be created that is $0$ if $A_1=A_1$ and $1$ if $A_1 = A_1^\d$.

  \texttt{data} $\gets$ \texttt{matrix}$(\{H_1, A_1, \Lambda_1 = 0\} \})$
  
  \texttt{data\_copy} $\gets$ \texttt{matrix}$(\{H_1, A_1 = A_1^\d, \Lambda_1 = 1\})$
  
  \texttt{dr\_data} $\gets$ \texttt{bind\_rows(data, data\_copy)}

\item Fit a logistic regression for $\Lambda_1$ conditional on $A_1=a_1$ and $H_1=h_1$. The odds ratio at a given $a_1$ and $h_1$ is the estimate of the density ratio at time 1, $\hat{\r_1}$.

  \texttt{r1\_fit} $\gets$ \texttt{glm}$(\text{outcome} = \Lambda_1, \text{predictors} = \{H_1, A_1\}, \text{data} = \texttt{dr\_data})$

  \texttt{r1\_hat} $\gets$  \texttt{predict}$(\text{fit} =   \texttt{r1\_fit}, \text{prediction = odds ratio}, \text{new data = }\texttt{data})$

\item Create a new column in the duplicated data set which contains values of $A_0$ in the first row of a duplicated pair. The second row in a pair should be modified so that $A_0 = A_0^\d$. A new column $\Lambda_0$ should be created that is $0$ if $A_0=A_0$ and $1$ if $A_0 = A_0^\d$.

\begin{figure}[H]
\begin{center}
    \includegraphics[width=8cm]{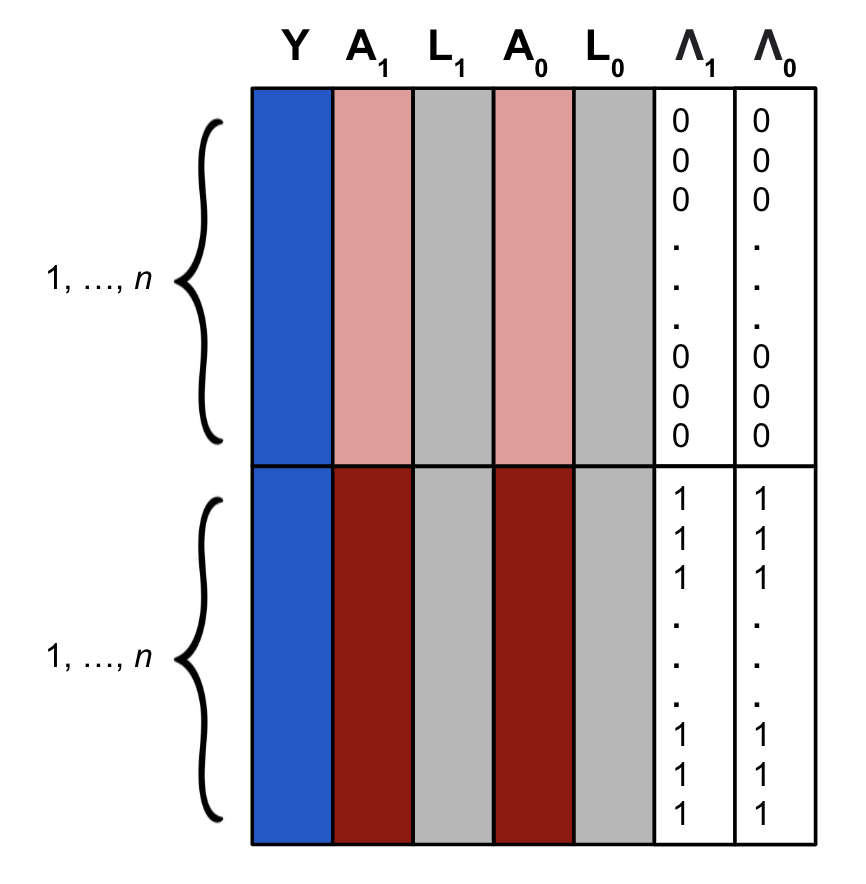}\label{illustration}
\caption{Illustration of the data used for density ratio estimation with the classification trick.}
\end{center}
\end{figure}

\item Fit a logistic regression for $\Lambda_0$ conditional on $A_0=a_0$ and $H_0=h_0$. The odds ratio at a given $a_0$ and $h_0$ is the estimate of the density ratio at time 0, $\hat{\r_0}$.

 \texttt{r0\_fit} $\gets$ \texttt{glm}$(\text{outcome} = \Lambda_0, \text{predictors} = \{H_0, A_0\}, \text{data} = \texttt{dr\_data})$

  \texttt{r0\_hat} $\gets$  \texttt{predict}$(\text{fit} =   \texttt{r0\_fit}, \text{prediction = odds ratio}, \text{new data = }\texttt{data})$

\item Multiply $\hat{\r_1}$ and $\hat{\r_0}$ together. This is the cumulative density ratio. Multiply the cumulative density ratio by $Y$. Compute the average. This is $\hat\E[\tilde{Y}_0]$.

\texttt{estimate} $\gets$ \texttt{mean(r1\_hat * r0\_hat *} $Y)$

\end{enumerate}

\subsubsection{Technical requirements for theoretical guarantees of estimators}

 The
  $\sqrt{n}$-consistency and asymptotic normality of the estimators
  proposed in \citet{diaz_nonparametric_2021}, and therefore the
  correctness of the confidence intervals output by the R software
  \emph{lmtp} \citep{nickpaper, lmtppkg}, relies on two important
  technical requirements. 
First, the function $\d$ must not depend on the distribution of the
data. Second, the exposure must be discrete or, if it is continuous,
$\d(\cdot, h_t, \epsilon_t)$ (as a
function of $a_t$) must be piecewise smooth invertible \citep{haneuse2013estimation,
  diaz_nonparametric_2021}. If any of these reequirements is violated,
important theoretical properties of the estimators such as
$\sqrt{n}$-consistent will fail. This means that important uncertainty
quantification measures such as confidence intervals, p-values, and
standard errors outputted by the package will be incorrect.

If the first requirement is violated, it may be possible to develop
$\sqrt{n}$-consistent estimators. For example,
\citet{kennedy2019nonparametric} developed non-parametric
$\sqrt{n}$-consistent estimators for the IPSI shown in Example \ref{exkennedy}, in
which $\d$ depends on the treatment mechanism. However, the IPSI
proposed by \citet{wen2021intervention} in Example \ref{exwen}, which does not
depend on the data distribution, can be estimated with the proposed
estimators. Here it is important to note that the the estimators for
the odds ratio IPSI of \citet{kennedy2019nonparametric} are not doubly
robust (and such estimators possibly cannot be constructed, since $\d$
depends on the propensity score), whereas the estimators of the
risk-ratio IPSI will be doubly robust.

If the second requirement is violated, it is not possible to construct
$\sqrt{n}$-consistent estimators. Intuitively, this is due to a lack
of ``smoothness'' in the parameter, and interested readers should
refer to literature on pathwise differentiability \citep[an accessible
review may be found
in][]{kennedy2022semiparametric}. 
In Example 
\textcolor{red}{3}, the static and dynamic interventions for corticosteroids
administration are possible to estimate using the proposed estimators
because the exposure is discrete. However, if the exposure were
continuous (e.g. a dosage), this would not be possible. The modified threshold function (Example 
\textcolor{red}{6}) also does not
meet the second requirement because $\d$ is not piecewise smooth
invertible. However, the modified shift functions (Example 
\textcolor{red}{7}) meet
both requirements and are estimable using the proposed estimators.

\subsection{Practical guidance and considerations}

 In practice we recommend using TMLE if the user is not concerned about model misspecification, because the estimates are guaranteed to remain within the observed outcome bounds. However, if model misspecification is a concern, SDR is significantly more robust, although there is a potential for estimates which fall outside the outcome bounds. An effective approach against model misspecification is to use the superlearner algorithm for estimating
the exposure and
outcome mechanisms. Superlearning
combines the predictions of multiple pre-specified statistical learning models via weighting to produce final estimates which are proven to perform as well as possible in large sample sizes 
\citep{vanderLaanPolleyHubbard07}.

To assess positivity violations, we recommend the researcher to visually inspect or examine summary statistics of the density ratios. Extremely high values of density ratios (the quantification of ``high'' being context-dependent) are akin to propensity score values at or near zero, generally indicate positivity violations, and may create unstable estimates. If violations are detected, the researcher may be interested in modifying the portion of the population receiving the intervention, or in cases of continuous exposures, making the intervened exposure level closer to the observed exposure level. Interventions should be designed to avoid both theoretical and practical positivity violations present in the data, the latter of which can potentially be caught in the exploratory data analysis stage. One option if the researcher cannot find a solution by study design to elimitate positivity violations is to truncate the density ratios as a certain threshold, e.g. the 98th quantile. This is akin to truncating a propensity score, and may have potential for biases \citep{leger2022causal}.

Analysts using the LMTP framework for time-varying or survival outcomes will also note that  discrete outcome intervals are required. The researcher must find a balance in choosing intervals that are both scientifically relevant (e.g. week-long intervals for a critically ill patient population would be too long) and in which there are still a number of outcomes which occur within the chosen interval. If the outcome is rare and the sample size is small, or if the sample size is very large and computationally intensive, the researcher may need to adjust the number of cross-fitting and/or superlearning cross-validation folds.

\subsection{Additional Application Details}

\subsubsection{Methods}

Baseline confounders include age, sex, race, ethnicity, body mass index (BMI), comorbidities
(cerebral vascular event, hypertension, diabetes mellitus, cirrhosis, chronic obstructive pulmonary disease, active
cancer, asthma, interstitial lung disease, chronic kidney disease, immunosuppression, HIV-infection, and home oxygen use), and hospital admission location. Time-dependent confounders include vital signs (heart rate, pulse oximetry percentage,
respiratory rate, temperature, systolic/diastolic blood pressure), laboratory results (blood urea nitrogen (BUN)-
creatinine ratio, creatinine, neutrophils, lymphocytes, platelets, bilirubin, blood glucose, D-dimers, C-reactive
protein, activated partial thromboplastin time, prothrombin time, arterial partial pressures of oxygen and carbon
dioxide), and concurrent pharmaceutical treatments. Concurrent treatments included vasopressors, diuretics, Angiotensin-converting enzyme
(ACE) and Angiotensin receptor blockers (ARBs), hydroxychloroquine, and tocilizumab.

In the case of multiple time-varying
confounders measured in one day, the clinically worst value was
used.
Although data was relatively complete due to manual abstraction
efforts, there were instances of patients missing laboratory
results at one or multiple time points. A combination of last observation carried forward and an
indicator for missing values was used to handle this informative missingness within the treatment and regression models \citep{burton2004missing}. 

The Superlearner ensemble algorithm utilized 5-fold cross-validation and candidate libraries included generalized linear models, multivariate adaptive regression splines \cite{earth}, random forests \cite{ranger}, and
extreme gradient boosted trees \cite{xgboost} An assumption was made that the time-varying
covariates from the previous two days (i.e. lag of 2 days) was
sufficient to capture the mechanisms to reflect the collection of
laboratory results at minimum 48-hour intervals. A 5-fold
cross-fitting component was implemented for the final estimator to
prevent variation in a certain sample split from biasing the final
estimate \citep{zivich2021machine}. All analyses were conducted in R Version 4.1.2 with the packages tidyverse \cite{tidyverse} for data cleaning and plotting, and lmtp \cite{lmtppkg} and SuperLearner \cite{superlearner} for estimation.

\subsubsection{Discussion}

 There are limitations specific to our illustrative example. While the estimated effect of a less aggressive intubation strategy can help to understand an underlying biological or mechanistic process, it may not provide clinical guidance if the treatment strategy changes over time. The estimates themselves depend on the natural value of treatment, and this is dependent on the state of clinical practice during the study time frame (Spring 2020). 
In addition, we cannot rule out unmeasured confounding in the
exposure, outcome, and loss to follow-up mechanisms. We also cannot be sure whether the informative right censoring is correctly specified, since patients were lost to follow-up for different reasons (e.g. discharge to home vs. assisted living).

\pagebreak


\bibliography{refs}










\doublespacing

\begin{abstract}
This tutorial discusses a recently developed methodology for causal
inference based on \textit{longitudinal modified treatment policies}
(LMTPs). LMTPs generalize many commonly used parameters for causal
inference including average treatment effects, and facilitate the
mathematical formalization, identification, and estimation of many
novel parameters. LMTPs apply to 
a wide variety of exposures, including binary, multivariate, and
continuous, as well as interventions that result in violations of the
positivity assumption. LMTPs can accommodate time-varying treatments
and confounders, competing risks, loss-to-follow-up, as well as
survival, binary, or continuous
outcomes. 
This tutorial aims to illustrate several practical uses of the LMTP framework, including describing different estimation
strategies and their corresponding advantages and disadvantages. We provide numerous examples of types of research questions which can
be answered within the proposed framework. We go into more depth with one of these examples---specifically, estimating the effect of delaying
intubation on critically ill COVID-19 patients'
mortality. We demonstrate the use of the open-source R package \emph{lmtp} to estimate the effects, and we provide code on \url{blinded for review}. 
\end{abstract}

\noindent \textbf{Keywords:} longitudinal modified treatment policies, static interventions, dynamic treatment rules, natural value of treatment, time-varying confounding, time-varying exposure, causal inference

\newpage

\section{Introduction}

Under a counterfactual framework, evaluating causal relations from data requires conceptualizing hypothetical modifications to a
causal agent of interest, and then assessing changes to an outcome
distribution under such hypothetical modifications.
Most of the causal
inference literature has focused on interventions that are defined
\textit{statically}, for example, answering questions about what would
have happened if all patients in the population received a certain
treatment course versus another. While this approach has spurred significant
progress, it is limited in multiple scenarios. For instance,
the definition of causal effects for continuous exposures such as the
duration of surgery \citep{haneuse2013estimation}
requires conceptualizing hypothetical modifications whereby surgery
time is reduced relative to the factually observed surgery time. Likewise, certain applications (e.g., in AIDS research
\citep{cain2010start}) consider varying the time-frame for treatment, which cannot be formalized in a static
intervention framework. Furthermore, the identification of causal
parameters based on static interventions is not possible in the
presence of violations of the positivity assumption, which roughly
states that the intervention considered must occur with positive probability
within all strata of confounders \citep{petersen2012diagnosing}. While
\textit{dynamic} interventions (which consider different interventions
according to strata of covariates) have also been developed, they do not readily offer a solution to the aforementioned challenging examples.

Longitudinal modified treatment policies (LMTP) are a novel
methodological development that offers a general
approach to translating complex research questions into mathematical
causal estimands, and have the potential to address several common
limitations of static and dynamic interventions. In brief, LMTPs
generalize static and dynamic interventions by allowing the
hypothetical intervention to depend on the \textit{natural value of
  treatment}, i.e. the value that treatment would take at time $t$ if an intervention was discontinued right before time $t$ \citep{YoungHernanRobins2014}. LMTPs accommodate a range of
interventions \citep{robins2004effects, taubman2009intervening, shpitser2012effects, munoz2012population,  haneuse2013estimation,
richardson2013single, YoungHernanRobins2014, diaz_nonparametric_2021} including binary,
categorical, continuous, and multiple exposures, and a range of outcomes including binary, continuous, or time-to-event outcomes with possible
competing risks and informative right-censoring, in both
point-in-time and time-varying settings. LMTP can also address violations of the positivity
assumption because it allows researchers to define interventions for
which positivity holds by design.
 
In this tutorial, we provide a guide to understanding and applying LMTP. We begin by introducing the different categories of
and modified. We then delve into the specifics of the LMTP framework,
covering point-in-time and time-varying exposures. We highlight
several estimation procedures 
and provide numerous
examples of research questions that can be addressed using the
framework. As a case study, we apply the LMTP framework to estimate
the effect of intubation timing on mortality in COVID-19 patients,
using a real-world longitudinal observational data set. We provide
detailed descriptions of the study design and analytical methods, as
well as code and synthetic data to facilitate replication by
future researchers.
 
\section{Notation and general setup}

Consider a sample of i.i.d. observations $Z_1, ..., Z_n$ drawn from a
distribution $\P$. This $\P$ represents a longitudinal process and may
contain any number of time points, but we will describe a distribution
with only two time points, $t \in \{0,1\}$, for simplicity. For each unit in the
study, we observe a set of random variables
$Z = (L_0, A_0, L_1, A_1, Y)$. At the first time point, baseline
covariates $L_0$ affect the baseline exposure, $A_0$. At the second
time point, we observe covariates $L_1$ and exposure $A_1$, which are
themselves affected by $L_0$ and $A_0$, and have the potential to
change from their respective baseline values (time-varying). An
outcome $Y$ is measured at the end of a defined follow up period. Each
endogenous variable $L_0, A_0, L_1, A_1,$ and $Y$ has a corresponding
exogenous variable $U$, representing the unmeasured, external factors
affecting each measured process. We may use the following simplified directed acyclic graph (DAG) \citep{pearl1998graphs} to denote the set-up.

\[\begin{tikzcd}[row sep=3.15em]
	& {A_0} && {A_1} \\
	{L_0} && {L_1} && Y
	\arrow[from=2-1, to=2-3]
	\arrow[from=1-2, to=2-3]
	\arrow[from=2-3, to=1-4]
	\arrow[from=1-4, to=2-5]
	\arrow[from=2-3, to=2-5]
	\arrow[from=1-2, to=1-4]
	\arrow[from=2-1, to=1-2]
	\arrow[from=1-2, to=2-5]
	\arrow[curve={height=24pt}, from=2-1, to=2-5]
\end{tikzcd}\]

We will use $H_t$ as a shorthand notation for the history of data
measured up to right before $A_t$. For example, $H_0=L_0$, and
$H_1 = (L_0, A_0, L_1)$. We conceptualize causal interventions, or treatment policies, in
terms of hypothetical interventions on nodes of the DAG
\citep{pearl2016causal}.
First, consider a user-given function $\d_0(a_0, h_0, \epsilon_0)$ which maps a treatment value $a_0$, a history $h_0$, and a randomizer $\epsilon_0$ into a potential treatment value. This randomizer $\epsilon_0$ adds stochasticity, but has a known distribution (we give examples below). The intervention at time $t=0$ is defined by removing node $A_0$ from the DAG and replacing it with $A_0^\d = \d_0(A_0, H_0, \epsilon_0)$.
This assignment generates counterfactual data $H_1(A_0^\d)$
and $A_1(A_0^\d)$, where the former is referred to as the
counterfactual history and the latter is referred to as the
\textit{natural value of treatment} \citep{robins2004effects,
  richardson2013single, YoungHernanRobins2014}, i.e., the value that
treatment would have taken if the intervention is performed at time
$t=0$ but discontinued thereafter.
At time $t=1$, the intervention is
likewise defined by a function $\d_1(a_1, h_1, \epsilon_1)$. However,
at $t=1$ (and all subsequent times if there are more than two time
points), the function must be applied 
applied to both the
natural value of treatment \emph{and} the counterfactual history. That is, at time $t=1$, the intervention
is
defined by removing node $A_1$ from the DAG and replacing it with
$A_1^\d = \d_1(A_1(A_0^\d), H_1(A_0^\d), \epsilon_1)$.

We refer to these longitudinal interventions, and the subsequent framework to identify and estimate effects under such interventions, as LMTPs. The main difference between a modified treatment policy (MTP) and a standard dynamic treatment rule is that in an MTP the intervention function $\d_t$ is allowed to depend on the natural value of treatment $A_1(A_0^\d)$.
We now give examples of how the functions $\d_t$ may
be defined, explain how they generalize static, dynamic, and
many stochastic interventions, and discuss novel and useful interventions
that may be defined using this setup.

\section{Illustrating the generality of LMTPs}

The function $\d_t(a_t, h_t, \epsilon_t)$ that defines the
intervention or treatment policy can be classified into four
categories of increasing generality: static, dynamic, stochastic, and
modified.  We summarize these hierarchical categories
in Table \ref{tab:interventions}.

\subsection{Static interventions}

In a static intervention, $\d_t(a_t, h_t, \epsilon_t)$ is a
constant value, i.e., it does not actually vary with $a_t$, $h_t$, nor
$\epsilon_t$.
\begin{example}[Average treatment effect]\label{exstatic}
  For the two time point example, one might examine the counterfactual
  outcomes in a hypothetical world in which all units are treated at
  both time points ($\d_t=1$ for $t\in\{0,1\}$), and contrast them to
  a hypothetical world in which no units are treated at either time
  point ($\d_t=0$ for $t\in\{0,1\}$), giving rise to the well-known
  average treatment effect.
\end{example}

\subsection{Dynamic interventions}
In a dynamic intervention, the function $\d_t$ assigns a treatment
value according to a unit's covariate history $h_t$, but
does not vary with $a_t$ nor
$\epsilon_t$. 
This is often used in observational studies when study units need to meet an indication of interest for a treatment to reasonably begin. In medical research, this indication could be a severity of illness indicator. In policy research, this indication could be a socioeconomic threshold which would trigger a policy of interest to apply. We provide two epidemiological examples of a dynamic intervention below.   
\begin{example}\label{eq:exhiv}[Antiretroviral initiation for HIV patients]
  One of the first uses of dynamic interventions was in the context of
  HIV, where investigators were interested in the effect of initiating
  antiretroviral therapy for a person with HIV if their CD4 count
  falls below a threshold, e.g. 200 cells/$\mu$l
  \citep{hernan2006comparison}. This can be described mathematically
  as \[\d_t(h_t)=\begin{cases}
      1 &\text{ if } l_t^*<200\\
      0&\text{ otherwise,}
    \end{cases}\]
    where $L_t^*$ is a variable in $H_t$ that denotes CD4 T-cell count.
\end{example}


\begin{example}\label{eq:excort}[Corticosteroids for COVID-19 hospitalized patients]
  A more recent example of a dynamic treatment regime application
  is to study the effect of initiating a corticosteroids
  regimen for COVID-19 patients. \citet{hoffman2022comparison} studied a hypothetical policy of initiating corticosteroids for six days if and when a COVID-19 patient met a severity of illness criteria (i.e. low levels of blood oxygen). In notation,
  
  

  \[\d_t(h_t)=\begin{cases}
      1 &\text{ if } l_s^*=1 \text{ for any } s\in\{t-5,\ldots, t\}\\
      0&\text{ otherwise,}
    \end{cases}\]
      where $L_t^*$ is a variable in $H_s$ that denotes low levels of blood oxygen.
 \end{example}

\subsection{Stochastic interventions}
Stochastic interventions allow the function
$\d_t(a_t, h_t, \epsilon_t)$ to vary with some user-given randomization
component $\epsilon_t$ and an individual unit's prior
history $h_t$, but not $a_t$.

\begin{example}[Stochastic pollution exposures based on population density]\label{exstoch}

We can imagine a stochastic intervention for a continuous exposure of pollution, for example particulate matter (PM2.5). An environmental health researcher might be interested in studying asthma hospitalization rates under a hypothetical world in which individuals had different exposure to PM2.5. Since it is generally difficult for urban areas to exhibit comparable pollution levels to rural areas, we may wish to study a stochastic intervention in which rural regions have a PM2.5 exposure level drawn from $\epsilon_1 \sim Normal(3, 1)$, and urban regions have a PM2.5 exposure level drawn from $\epsilon_2 \sim Normal(7, 2)$.

 \[\d(\epsilon_t, h_t)=\begin{cases}
      \epsilon_1 &\text{ if } l_t^* = 1,\\
      \epsilon_2 &\text{ otherwise.}
    \end{cases}\]
where $l_t^*$ is an indicator of urban vs not. 
\end{example}

\subsection{Modified treatment policies}
Lastly, the more general type of intervention that we discuss is a
modified treatment policy, in which the intervention function
$\d_t(a_t, h_t, \epsilon_t)$ is allowed to depend on some combination of $a_t, h_t, \epsilon_t$. 

\begin{example}\label{exsmoking}
  One of the earliest proposals of an MTP was
  \citet{robins2004effects}'s proposal to study a hypothetical policy in
  which half of all current smokers quit smoking forever. This intervention is motivated by the infeasibility of studying a world in which all current smokers quit smoking forever, since genetics, environment, and many other factors (likely unmeasured) will always create some portion of current smokers who will never quit. Letting
  $A_t$ denote a random variable denoting smoking, this intervention
  may be represented in notation as
  \[\d_t(a_t,\epsilon_t)=\begin{cases}
      0 & \text{ if } \epsilon_t<0.5 \text{ and } a_t=1\\
      a_t & \text{ otherwise, }
    \end{cases}\] where $\epsilon_t$ is a random draw from a uniform
  distribution in $(0,1)$.  
\end{example}

\begin{example}[Threshold intervention]\label{exthreshold}
In a threshold function, all natural exposure values which fall outside of a certain boundary are intervened upon to meet a threshold. This type of intervention was proposed by \citet{taubman2009intervening} to assess the effect of multiple lifestyle interventions (e.g. exercising at least 30 minutes a day, maintaining a BMI < 25, and consuming at least 5g of alcohol a day) on
the risk of coronary heart disease. We can consider one such intervention on BMI in notation as,

\[\d_t(a_t)=\begin{cases}
      a_t & \text{ if } a_t < 25\\
      25 & \text{ otherwise. }
    \end{cases}\]
    
\end{example}

\begin{example}[Multiplicative or additive shift functions]\label{exshift}

  Shift functions assign treatment by modifying the natural value of
  the exposure by some constant $\delta$. This intervention can be
additive onto the exposure value, such as \citet{haneuse2013estimation}'s estimates of the effect of a hypothetical intervention to reduce lung cancer resection surgeries lasting longer than 60 minutes by 15 minutes.

    \[\d_t(a_t)=\begin{cases}
      a_t & \text{ if } a_t \ge 60 \\
      a_t - 15 & \text{ otherwise. }
    \end{cases}\] 
Additive shift interventions were originally proposed by \citet{munoz2012population} and applied to shifts in leisure-time physical activity.
\noindent This shift function could also change the exposure on a
  multiplicative scale.  For example, we
  may be interested in studying the effect of an intervention which
  doubles the number of street lights for roads with less than 10
  lights per mile on nighttime automobile accidents.

  \[\d_t(a_t)=\begin{cases}
      a_t & \text{ if } a_t \ge 10 \\
      2a_t & \text{ otherwise. }
    \end{cases}\]

\end{example}

\noindent We provide additional examples of interesting incremental propensity score interventions in the Appendix.

\section{Causal estimands and identifying parameter}

Once an intervention is specified, the counterfactual outcomes of
observations under a specific $\d=(\d_0, \d_1)$ are denoted as
$Y(\d)$. Causal effects are defined as contrasts between the
distributions of $Y(\d)$ under different interventions, $\d'$ and
$\d^\star$. In this tutorial, we focus on
$\E[Y(\d') - Y(\d ^ {\star})]$ as our causal estimand of interest. The
functions $\d'$ and $\d^\star$ may be any combination of static,
dynamic, stochastic, or modified
interventions.

The next step in a formal causal inference analysis is to write the
counterfactual expectation $\E[Y(\d)]$ as a formula that depends only
on the observed data distribution---i.e., an
identifying formula. This will generally require assumptions, some of
which are untestable. The mathematically rigorous form of the
assumptions is given elsewhere \citep{YoungHernanRobins2014,
  diaz_nonparametric_2021}, but we state them here in simple terms:

\begin{assumption}[Positivity or common support \citep{YoungHernanRobins2014}]
  If it is possible to find an observation with history $h_t$ with an
  exposure of $a_t$, then it is also possible to find an observation
  with history $h_t$ with an exposure $\d(a_t, h_t, \epsilon_t)$.
\end{assumption}
\begin{assumption}[Strong sequential randomization]
  This assumption states
  that 
  the exposure is independent of all future counterfactual data
  conditional on the measured history at every time point. This is
  generally satisfied if $H_t$ contains all common causes of $A_t$ and
  $(L_{t+1}, A_{t+1}, \ldots, L_\tau, A_\tau,\allowbreak Y)$, where $\tau$ is
  the last time point in the study.
\end{assumption}

\begin{assumption}[Weak sequential randomization]
  This assumption states
  that 
  the exposure is independent of all future covariate data
  conditional on the measured confounders at every time point. This is
  generally satisfied if $H_t$ contains all common causes of $A_t$ and
  $(L_{t+1}, \ldots, L_\tau, \allowbreak Y)$, where $\tau$ is
  the last time point in the study.
\end{assumption}

The strong and weak versions of the sequential randomization are different in that the former requires unconfoundnedness of $A_t$ with all future $A_s:s>t$ conditional on $H_t$, whereas the latter does not. Identification of interventions that depend on the natural value of treatment (such as Examples \ref{exsmoking}-\ref{exshift} above) require the strong version of sequential randomization. Interventions that do not depend on the natural value of treatment (such as Examples \ref{exstatic}-\ref{exstoch}) require the weak version of sequential randomization. 

\subsection{Positivity}

By design, MTPs may help meet the positivity assumption, since the function $\d$ is given by the user. Violations to positivity can be structural, meaning there are certain characteristics of an individual or unit which will never yield receipt of the treatment assignment under the intervention. This type of positivity violation will not improve even with an infinite sample size. Violations to positivity can also be practical, meaning due to random chance or small datasets, there are certain covariate combinations with zero or near-zero predicted probabilities of treatment. For time-varying treatments, positivity must be maintained at each time point, which can increase the likelihood of positivity assumption violations \citep{van2018targeted}. Positivity violations can increase the bias and variance of estimates and severely threaten the validity of casual inference analyses when not addressed \citep{petersen2012diagnosing}. The LMTP framework's flexibility allows researchers to formalize scientific questions with a higher chance of meeting the positivity assumption \citep{diaz_nonparametric_2021}.

This can be seen in the MTPs described above, for instance, the additive shift in Example \ref{exshift}. It is intuitive to conceptualize a world in which a continuous exposure is instead observed at some fixed value higher or lower than it was factually observed for every unit in the study; for example, imagine if surgery times were 15 minutes shorter for all lung resection biopsies. However, this type of uniform hypothetical modification is destined for structural positivity violations, because at the lowest end of the observed exposure range, there will by definition be no support for the intervened exposure level $\d(a_t)$ (much less conditional on the observation's history $h_t$). This can be avoided by constraining the range of $a_t$ affected by the hypothetical intervention, so that no $\d(a_t)$ values are produced outside the
observed range of
$A$. The intervention function can also be modified to accommodate any other remaining structural positivity violations. For example, clinical knowledge may inform us that a treatment of interest will never be administered after a certain amount of time since a disease diagnosis has passed, so the hypothetical intervention would restrict the values of $t$ in which the intervention can occur.

\subsection{Identification formula}

Under Assumptions 1 and 2, or 1 and 3, the estimand is identified by the
generalized g-formula \citep{Robins86}. A re-expression of this generalized g-formula \citep{bang2005doubly, diaz_nonparametric_2021}  involves recursively defining the expected outcome under the
intervention, conditional on the observation's observed exposure and
history, beginning at the final time point, and proceeding until the
earliest time point. We illustrate the
g-formula for two time points below:

\begin{enumerate}

\item Start with the conditional expectation of the outcome $Y$ given $A_1=a_1$ and $H_1=h_1$. Let this function be denoted $Q_1(a_1, h_1)$.
\item Evaluate the above conditional 
  expectation of $Y$ if $A_1$ were changed to $A^\d_1$, which results in 
  a pseudo outcome $\tilde Y_1=Q_1(A^\d_1, H_1)$.
\item Let the true expectation of $\tilde Y_1$ conditional on
  $A_0=a_0$ and $H_0=h_0$ be denoted $Q_0(a_0, h_0)$.
\item Evaluate the above
  expectation of $\tilde Y_1$ if $A_0$ were changed to $A^\d_0$, which results in
  $\tilde Y_0=Q_0(A^\d_0, H_0)$.
\item Under the identifying assumptions, we have
  $\E[Y(\d)]=\E[\tilde Y_0]$.
  
\end{enumerate}

\section{Estimation}

Once a causal estimand is defined and identified, the researcher's task is to estimate the statistical quantity, e.g. $\E[\Tilde{Y}_0]$. We
now review several estimators, both parametric and non-parametric.

\subsection{Parametric estimation}

 The simplest option for estimation is to fit parametric outcome regressions for each step of the g-formula identification result. This ``plug-in'' esimator is often referred to as the parametric g-formula or g-computation. 
 Another option is to use an estimator which relies on the exposure mechanism, for example, the Inverse Probability Weighting (IPW) estimator. IPW estimation involves reweighting the observed outcome by a quantity which represents the likelihood the intervention was received, conditional on covariates. 
 We provide algorithms for G-computation and IPW estimation for two time points in the Appendix.

Obtaining point estimations with the g-computation and IPW algorithms is computationally straightforward. If the exposure regression for IPW or outcome regression for
g-computation are estimated using pre-specified parametric statistical models, standard errors for the estimate can be computed using bootstrapping or the Delta method. However, in causal models with large numbers of covariates and/or complex mathematical relations between confounders, exposures, and
outcomes, parametric models are hard to pre-specify, and they are unlikely to consistently estimate the regressions. If the regression for the outcome (for g-computation) or treatment (for IPW) are misspecified, the final estimates will be biased.

One way to alleviate model misspecification is to use flexible approaches which incorporate model selection (e.g. 
LASSO, splines, boosting, random forests, ensembles thereof, etc.) to estimate the exposure or outcome
regressions. Unfortunately, there is generally not statistical theory to support the standard errors of the g-computation or IPW estimators with such data-adaptive regressions. Standard inferential tools such as the bootstrap will fail because these estimators generally do not have an asymptotically normal distribution after using data-adaptive regressions \citep{van2011targeted}. Thus, other methods are needed to accommodate both model selection and flexible regression techniques while still allowing for statistical inference.

\subsection{Non-parametric estimation}

  \citet{diaz_nonparametric_2021} proposed general non-parametric estimators for LMTPs. These estimators use both an exposure and outcome regression, and allow the use of machine learning to estimate the regressions while still obtaining valid statistical uncertainty quantification on the final estimates. 
  The estimators also have double robustness properties. 

 The two non-parametric estimators proposed in \citet{diaz_nonparametric_2021} and encoded in the R package \emph{lmtp} \citep{lmtppkg, nickpaper} are Targeted Maximimum Likelihood Estimation (TMLE) \citep{van2011targeted, van2012targeted, diaz_nonparametric_2021} and Sequentially Doubly Robust (SDR) estimation \citep{luedtke2017sequential, rotnitzky2017multiply, diaz_nonparametric_2021}. A third non-parametric estimator, iterative TMLE (iTMLE), is not encoded in the R package but could be adapted from \citet{luedtke2017sequential}.  TMLE is a doubly robust estimator for a time-varying treatment in the sense that is consistent as long as all outcome regressions for times $t>s$ are consistently estimated, and all treatment mechanisms for times $t\leq s$ are consistent, for some time $s$. In contrast, SDR and iTMLE are sequentially doubly robust in that they are consistent if for all times $t$, either the outcome or the treatment mechanism are consistently estimated \citep{luedtke2017sequential, diaz_nonparametric_2021}. Since TMLE and iTMLE are substitution estimators, they are guaranteed to produce estimates which remain within the observed outcome range. SDR and iTMLE produce estimators with more robustness than TMLE. Table \ref{tab:est-props} compares the characteristics of various
estimators, and we provide practical guidance for choosing between estimation techniques in the Appendix.

\section{LMTP applications in practice}

The LMTP framework enables researchers to convert many queries with
complex exposures into causal estimands that can be estimated using
available analytical tools.
In the worked example, we estimate the effect of delaying intubation
on mortality. Other examples of LMTP applications in similar
populations include studying the effect of a delay in intubation on an
outcome of acute kidney injury, where death is a competing risk
\citep{diaz2022causal}, or studying an intervention on a continuous
measure of hypoxia in acute respiratory distress patients
\citep{diaz_nonparametric_2021}. LMTP also accommodates interventions
involving multiple treatments, such as delaying intubation by one day
\emph{and} increasing fluid
intake. 

There are multiple other examples of researchers using LMTP to answer real-world problems. \citet{nugent2021evaluating} used an LMTP to study the effects of mobility on COVID-19 case rates. Specifically, they studied the effect of a longitudinal modified shift on the observed mobility distribution on the number of newly reported cases per 100,000 residents two weeks ahead. A similar analysis could be done to look at interventions on masking or vaccination policies. Mobility, masking, and vaccination are important examples of when static or dynamic policy estimands may be unappealing because of the geographical, political, and cultural variation that exists even within relatively small regions. 
This applies to environmental exposures and health policies as well. For instance, \citet{rudolph2022effects} and \citet{floriana} used LMTP frameworks to estimate the effects of Naxolone access laws and disability denial rates, respectively, on opioid overdose rates. 

Other examples of LMTP applications include \citet{jafarzadeh2022relationship}'s study of the effect of an intervention on knee pain scores over time on an outcome of knee replacement surgery. \citet{huling2022public} investigated the effects of interventions to public health nursing on the behaviors of clients in the Colorado Nurse Support Program. \citet{mehta2021primary} researched whether an increase in primary care physicians has an effect on post-operative outcomes in patients undergoing elective total joint replacement. 
Tables \ref{tab:bin-table} and \ref{tab:cont-table} show additional application examples of static, dynamic, stochastic, and modified interventions.

\section{Illustrative example}

\subsection{Motivation}

In the following application, we study a clinical question, ``what is the effect of delaying intubation of invasive mechanical ventilation (IMV) on mortality for hospitalized COVID-19 patients in New York City's first COVID-19 wave?" The relevance of this clinical question is justified in \citet{diaz2022causal}.
Studying the effect of IMV is particularly ill-suited for static interventions because there is no scenario in which the initiation of IMV at a certain study time could be uniformly applied across all critically ill patients. 
A dynamic intervention, which could help to evaluate a world in which patients are intubated when they meet a certain severity threshold, e.g. an oxygen saturation breakpoint, is also not clinically relevant because initiation of IMV may vary considerably between providers dependent on training, hospital policies, and ventilator availability \citep{tobin2020caution, perkins2020recovery}. For this reason, an LMTP which varies the natural time of intubation by only a minimal amount (for example, one day), may be a more realistic hypothetical intervention to study when considering the mechanistic effect of a delay-in-intubation strategy on mortality.
 
\subsection{Measures and analysis}

The population of interest is adult patients hospitalized and diagnosed with COVID-19 during Spring 2020.  The cohort contains 3,059 patients who were admitted to NewYork-Presbyterian's Cornell, Queens, and Lower Manhattan locations between March 1-May 15, 2020 \citep{goyal2020clinical, schenck2021critical, diaz2022causal}. This research was approved with a waiver of informed consent by the Weill Cornell Medicine Institutional Review Board (IRB) 20-04021909.

The exposure of interest is the maximum level of daily supplemental oxygen support. This can take three categories, 0: no supplemental oxygen, 1: non-IMV supplemental oxygen support, and 2: IMV. The intervention of interest describes a hypothetical world in which patients who require IMV have their intubation delayed by one day, and instead receive non-IMV supplemental oxygen support on the observed day of intubation.

\begin{equation*}\label{eq:exmtp}
\d_t(a_t,h_t) =
  \begin{cases}
    1 &\text{ if } a_t=2 \text{ and } a_s \leq 1 \text{ for all } s < t,\\
   a_t & \text{ otherwise}
 \end{cases}
\end{equation*}

\noindent Since patients in this observational data are subject to loss to follow up via discharge or transfer, the intervention additionally includes observing patients through all 14 days from hospitalization. 
    
The primary outcome is time to death within 14 days from hospitalization. 
 Hospital discharge or transfer to an external hospital system is considered an informative loss to follow-up. The causal estimand of interest is the difference in 14-day mortality rates between a hypothetical world in which there was a one-day delay in intubation and no loss to follow-up, and a hypothetical world in which there was no loss to follow-up and no delay in intubation.

Since the intervention is an MTP, we require positivity and strong sequential randomization to identify our parameter of interest (Identifying Assumptions 1 and 2). The common causes we assume to satisfy the latter requirement include 37 baseline confounders and
14 time-varying confounders per time point.

The R package \emph{lmtp} \citep{lmtppkg, nickpaper} was used to
obtain estimates of the difference in 14-day mortality rates under the
two proposed interventions using SDR estimation. A 
superlearner \citep{Wolpert1992, Breiman1996,
  vanderLaanPolleyHubbard07} library of various candidate learners was
used to estimate the intervention and outcome estimation. We demonstrate code on synthetic data 
at \url{blinded for review}  
  and provide several additional details in the Appendix. 

\subsection{Results}

\begin{figure}[H]
\caption{Panel A: Estimated incidence of mortality between a Delayed Intubation MTP (blue) and No intervention (red). Panel B: Estimated incidence difference in mortality if the Delayed Intubation MTP were implemented during Spring 2020. In both panels, 95\% simultaneous confidence bands \citep{westling2020correcting} cover the sets of point estimates.}
\centering
\includegraphics[width=17cm]{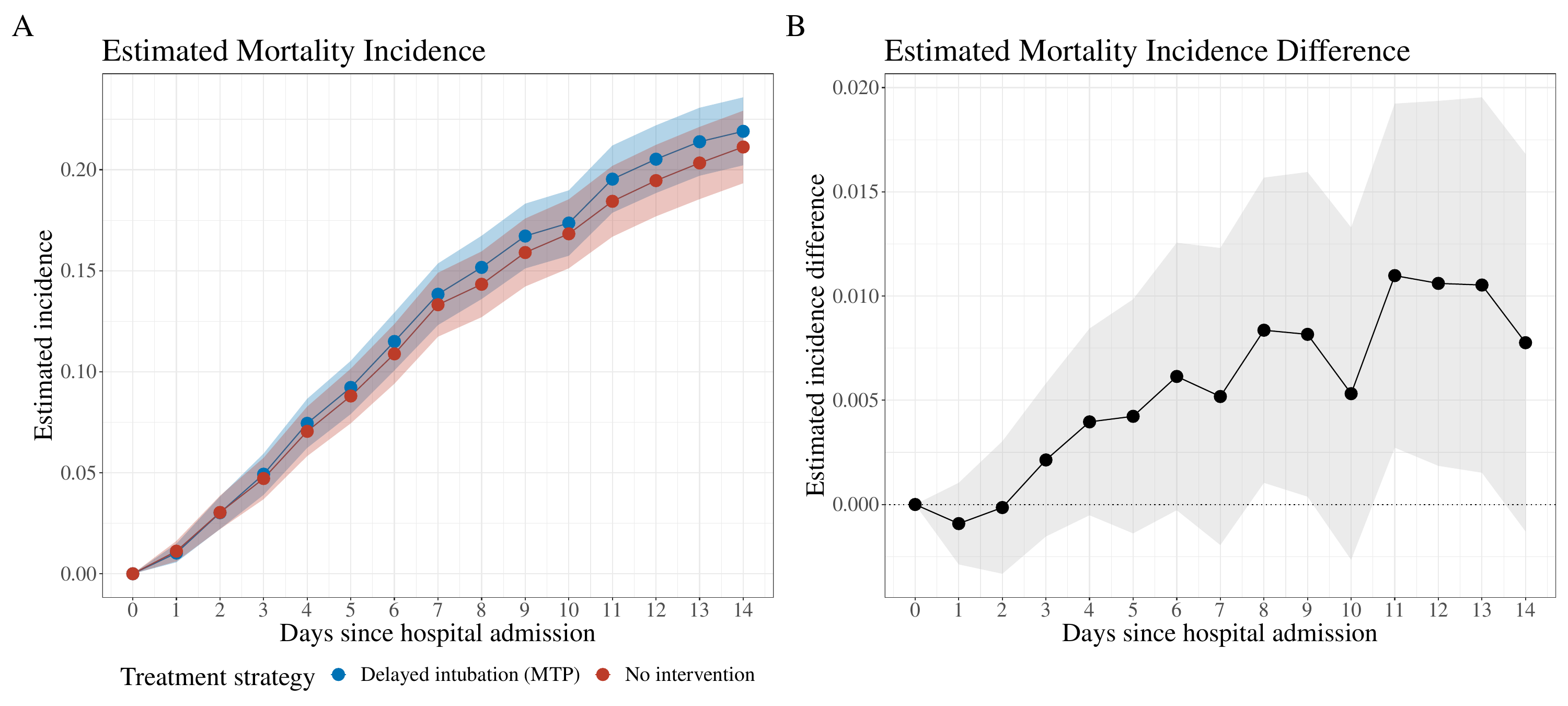}\label{fig1}
\end{figure}

The estimated 14-day mortality incidence under no intervention on intubation was 0.211 (95\% CI 0.193-0.229). The same incidence under a hypothetical LMTP in which intubation were delayed by 1 day was
0.219 (0.202-0.236). The estimated incidences across all time points are shown in Figure \ref{fig1}. 

\section{Discussion}

The LMTP
framework provides a
unified and comprehensive approach for defining,
identifying, and estimating relevant causal parameters, including those that involve challenges such as loss-to-follow-up, survival analysis,
missing exposures, competing risks, and 
interventions that include multiple exposures and/or continuous exposures. These causal parameters may be formulated to reduce positivity violations. In
addition, the existing packages that implement doubly and sequentially robust estimators
enable researchers to take advantage of statistical learning
algorithms to estimate the intervention and outcome mechanisms, thereby increasing the likelihood of estimator consistency.

 
While the LMTP framework expands the applied researcher's toolbox, there are considerations and
limitations to its implementation in real-world applications. First,
using it in a longitudinal setting requires discretizing time
over intervals. Depending on the data collection process, this may
cause issues in temporality or loss of data granularity. This discretization may also cause issues in small sample sizes if there are very few outcomes within a certain time point and the applied researcher hopes to estimate the outcome regressions using any statistical learning algorithm which segments the data for training/testing. Second,
although the formulation of MTPs may alleviate positivity
violations, some violations may still occur. 
Possible solutions, such as
truncating the density ratios at a certain threshold, are arbitrary and the potential for bias is unclear. Third, some particular estimator applications may be computationally intensive.  
Despite these limitations, we hope
the overview of LMTPs, illustrative example, and corresponding Github
repository are a useful toolset for researchers hoping to implement
LMTPs into their applied work.

\newpage

 \section*{Tables}
 
\begin{table}[H]
\begin{threeparttable}
\caption{An overview of common intervention types in the causal inference literature and whether they are estimable using the \emph{lmtp} R package.}
\centering
\begin{tabular}{  m{11cm} m{4.5cm} } 
\toprule
  \textbf{Intervention and definition} & 
\textbf{Estimable with \emph{lmtp}?} \\
  \midrule
  \textbf{Static:} all units receive the same treatment assignment  & Yes\tnote{a} \\ 
  \midrule
  \textbf{Dynamic:} A unit’s treatment assignment is determined according to their prior exposure and/or covariate history
 & Yes\tnote{a}  \\ 
   \midrule
   \textbf{Stochastic:} A unit’s treatment assignment is determined via a randomization component, and possibly their prior exposure or covariate history & See assumptions\tnote{b}  \\ 
    \midrule
  \textbf{Modified:} A unit’s treatment assignment is determined according to their natural exposure value, and possibly their prior exposure or covariate history, and possibly with a randomization component & See assumptions\tnote{b}  \\ 
  \hline
\end{tabular}
\begin{tablenotes}
 \item [a] The software should only be used with discrete exposures. The software will output a result with a continuous exposure, but this estimator will not have good statistical properties.
     \item [b]Assumptions require the intervention function $\d$ does not depend on the distribution $\P$, and,
one of: (1) the exposure is discrete or (2) the exposure is continuous but $\d$ satisfies piecewise smooth invertibility. This includes many, but not all, modified and stochastic interventions; see Technical Requirements in Appendix for more details.
\end{tablenotes}
\label{tab:interventions}
\end{threeparttable}
\end{table}

\begin{table}[H]
\begin{threeparttable}
\caption{Comparison of statistical properties for five possible estimators for LMTPs: g-computation (G-COMP), inverse probability weighting (IPW), targeted maximum likelihood estimation (TMLE), sequentially doubly robust estimation (SDR), and iterative TMLE (iTMLE).}
\centering

\begin{tabular}[t]{l c c c c c}
\toprule
\textbf{Statistical Property} & \textbf{G-COMP} & \textbf{IPW} & \textbf{TMLE} & \textbf{SDR} & \textbf{iTMLE}\tnote{a} \\
\midrule
Uses outcome regression & \usym{1F5F8} &  & \usym{1F5F8} & \usym{1F5F8} & \usym{1F5F8}\\
\midrule
Uses treatment regression &  & \usym{1F5F8} & \usym{1F5F8} & \usym{1F5F8} & \usym{1F5F8} \\
\midrule
Doubly robust\tnote{b}  &  &  & \usym{1F5F8} & \usym{1F5F8} & \usym{1F5F8}\\
\midrule
Sequentially doubly robust\tnote{c}  &  &  &  & \usym{1F5F8} & \usym{1F5F8}\\
\midrule
\makecell{Valid inference\tnote{d}\\ using parametric regressions\\ (i.e. generalized linear models)} & \usym{1F5F8} & \usym{1F5F8} & \usym{1F5F8} & \usym{1F5F8} & \usym{1F5F8} \\
\midrule
\makecell{Valid inference\tnote{d}\\ using data-adaptive regressions\\ (i.e. machine learning)} &  &  & \usym{1F5F8} & \usym{1F5F8} & \usym{1F5F8} \\
\midrule
\makecell{Guaranteed to stay within\\ observed outcome range} & \usym{1F5F8} &  & \usym{1F5F8} &   & \usym{1F5F8} \\
\bottomrule
\end{tabular}
\begin{tablenotes}
     \item [a] The iTMLE estimator is not currently available within the R package lmtp.
     \item [b] The estimator is consistent as long as all outcome regressions for times $t>s$ are consistently estimated, and all treatment mechanisms for times $t\leq s$ are consistent, for some time $s$.
     
     \item [c] The estimator is consistent if, for every time point, either the outcome regression or the treatment mechanism is consistently estimated.
     
     \item [d] Includes standard errors, confidence intervals, and p-values.
\end{tablenotes}
\label{tab:est-props}
\end{threeparttable}
\end{table}

\newpage

\begin{table}[H]
\begin{threeparttable}
\small
\caption{Examples of static, dynamic, stochastic, and modified interventions for (1) binary and (2) continuous point-in-time exposures defined via the generalized LMTP framework. These can be expanded to any categorical exposure. We denote a random variable drawn from Bernoulli distribution with probability $0.5$ with $\epsilon$ unless otherwise noted.}
\centering

\begin{tabular}[t]{p{2.5cm}p{2cm}p{4cm}p{6cm}}
\toprule \thead{Exposure} & \thead{Intervention} & \thead{``What if...''} & 
 \thead{Shift Notation}\\ 
\midrule

\multirow{2}{*}{\makecell{Point-in-time \\ Binary, \\ e.g. vaping \\ $(a=1)$}} & Static & \makecell{no one vapes} & 
    $\d(a, h, \epsilon) = 0$ \\

  \cmidrule{2-4}
   
    & Dynamic  & \makecell{only those working \\ non-standard work \\ hours $(l^*=1)$ vape} &  \makecell{$\d(a, h, \epsilon)=\begin{cases} 1  &\text{ if } l^*=1 \\ 0  &\text{ otherwise} \end{cases}$} \\

     \cmidrule{2-4}

  & Stochastic & \makecell{only a random half \\ of those working \\ non-standard work \\ hours $(l^*=1)$ vape} &  \makecell{$\d(a, h, \epsilon) = \begin{cases}  \epsilon &\text{ if } l^*=1 \\ 0 &\text{ otherwise} \end{cases}$} \\

  \cmidrule{2-4}

    & Modified & \makecell{a random half of \\ current vapers stop \\ vaping} & \makecell{$\d(a, h, \epsilon) = \begin{cases} \epsilon &\text{ if } a=1\\
        a &\text{ otherwise } \end{cases}$} \\
  
\midrule

\multirow{1}{*}{\parbox{3cm}{\raggedright Point-in-time Continuous, \\ e.g. exposure \\ to pollution as \\ measured by \\ the Air Quality \\ Index (AQI)  \\scale}} & Static & \makecell{all counties are \\ exposed to an AQI \\ of 10} & \makecell{$\d(a, h, \epsilon) = 10$} \\

   \cmidrule{2-4}

    & Dynamic  & \makecell{all urban ($l^*=1$) \\ counties are exposed \\ to an AQI of 40 and  \\ all rural ($l^*=0$) \\ counties are exposed to \\ an AQI of 20} & \makecell{$\d(a, h, \epsilon) = \begin{cases}  40 &\text{ if } l^*=1 \\
        20 &\text{ otherwise} \end{cases}$} \\

         \cmidrule{2-4}

& Stochastic & \makecell{all urban ($l^*=1$)  \\ counties are  exposed \\ to an AQI of \\ $\epsilon_1 \sim N(40,5)$ \\ and all rural ($l^*=0$) \\ counties are exposed \\ to an AQI of \\ $\epsilon_0 \sim N(20,5)$} &  \makecell{$\d(a, h, \epsilon) = \begin{cases}  \epsilon_1 &\text{ if } l^*=1 \\
        \epsilon_0 &\text{ otherwise} \end{cases}$} \\

  \cmidrule{2-4}

    & Modified & \makecell{all counties with an \\ AQI higher than 20 are \\ exposed to an AQI \\ 10\% lower than what \\ they were naturally \\ exposed to} & \makecell{$\d(a, h, \epsilon) = \begin{cases} a \times 0.9 &\text{ if } a > 20 \\ a &\text{ otherwise} \end{cases}$} \\

\bottomrule
\end{tabular}
\begin{tablenotes}
\end{tablenotes}
\label{tab:bin-table}
\end{threeparttable}
\end{table}

\newpage

\begin{table}[H]
\begin{threeparttable}
\small
\caption{Examples of static, dynamic, stochastic, and modified interventions for a binary time-varying exposure defined via the generalized LMTP framework. We index study time at $t=0$ and denote a random variable drawn from Bernoulli distribution with probability $0.5$ with $\epsilon_t$.}
\centering

\begin{tabular}[t]{p{2.5cm}p{2cm}p{3cm}p{8.5cm}}
\toprule \thead{Exposure} & \thead{Intervention} & \thead{``What if...''} &
 \thead{Shift Notation}\\ 
\midrule

\multirow{2}{*}{\makecell{Time-varying \\ Binary, e.g. \\ corticosteroids \\ 
 receipt $(a_t=1)$}} & Static & \makecell{all patients receive \\ corticosteroids for \\ the first 6 days of \\ hospitalization} & \makecell{$\d_t(a_t,h_t,\epsilon_t) = \begin{cases} 1 &\text{ if } t \leq 5 \\ 
    0 &\text{ otherwise} \end{cases} $} \\

  \cmidrule{2-4}

   & Dynamic  & \makecell{patients receive \\ corticosteroids \\ for 6 days once \\ they become \\ hypoxic $(l^*_t=1)$} &
      \makecell{$\d_t(a_t,h_t,\epsilon_t) = \begin{cases} 1 &\text{ if } l^*_s=1 \text{ for } s\in\{t-5,\ldots, t\} \\ 
    0 &\text{ otherwise} \end{cases} $} \\

        \cmidrule{2-4}

  & Stochastic & \makecell{a random half of \\ hypoxic $(l^*_t=1)$ \\ patients are given \\ corticosteroids \\ for 6 days} & \makecell{$\d_t(a_t,h_t,\epsilon_t) = \begin{cases} \epsilon_t &\text{ if } l^*_s=1 \text{ for } s\in\{t-5,\ldots, t\} \\ 
    0 &\text{ otherwise} \end{cases} $} \\
  
  \cmidrule{2-4}

  & Modified &
 \makecell{patients' receipt of \\ corticosteroids is \\ delayed by 1 day} & 
 \makecell{$\d_t(a_t,h_t,\epsilon_t) = \begin{cases} 0 &\text{ if } a_t=1 \text{ and } a_{t-1}=0\\ 
    a_t &\text{ otherwise} \end{cases} $} \\

\bottomrule
\end{tabular}
\begin{tablenotes}
\end{tablenotes}
\label{tab:cont-table}
\end{threeparttable}
\end{table}

\newpage

\bibliography{refs}


\section{Appendix}\label{appendix}

\subsection{Additional Intervention Examples}

\subsubsection{Dynamic interventions}

\begin{example}\label{eq:exhiv}[Dynamic antiretroviral initiation for HIV patients]
  One of the first uses of dynamic interventions was in the context of
  HIV, where investigators were interested in the effect of initiating
  antiretroviral therapy for a person with HIV if their CD4 count
  falls below a threshold, e.g. 200 cells/$\mu$l
  \citep{hernan2006comparison}. This can be described mathematically
  as \[\d_t(h_t)=\begin{cases}
      1 &\text{ if } l_t^*<200\\
      0&\text{ otherwise,}
    \end{cases}\]
    where $L_t^*$ is a variable in $H_t$ that denotes CD4 T-cell count.
\end{example}


\subsubsection{Stochastic interventions}

An additional category of interventions can be defined as functions which vary with some user-given randomizer $\epsilon_t$ (and possibly an individual unit's prior
history $h_t$) but not $a_t$. We will call these stochastic interventions. There are many real world examples of stochastic interventions, such as the lottery treatments of military drafts \citep{angrist1990lifetime}, twin births \citep{mcgue2010causal}, and college roommate assignments \citep{sacerdote2001peer}. Here we will focus on two stochastic interventions as they relate to the natural value of treatment. Of note, we do not consider regimes where $\epsilon$ has a 
distribution that depends on $\P$.





\begin{example}[Incremental propensity score interventions based on
  the odds ratio]\label{exkennedy}
  A recently developed type of stochastic interventions is
  \citet{kennedy2019nonparametric}'s incremental propensity score
  interventions (IPSI), which shifts a unit's probability of receiving
  treatment conditional on their according to some constant
  $\delta$. First, a shifted propensity score is defined:
\begin{equation*}
  \pi^\d_t(h_t) = \frac{\delta\pi_t(h_t)}{\delta\pi_t(h_t) + 1 - \pi_t(h_t)}
\end{equation*}
Then, a random variable $\epsilon_t$ is drawn from a uniform
distribution in $(0,1)$ and the intervention is defined as:
\begin{equation*}
\d_t(h_t, \epsilon_t, \pi_t) = I(\epsilon_t < \pi^\d_t(h_t)),
\end{equation*}
This IPSI is said to be based on an odds ratio because the odds ratio
of $\pi^\d_t(h_t)$ to $\pi_t(h_t)$ is equal to
$\delta$. \citet{naimi2021incremental} used an IPSI to study the
causal relationship between vegetable density consumption and the risk
of preeclampsia among pregnant women. They studied whether
preeclampsia would increase if women's propensity of eating a minimum
amount of vegetables increased by an odds ratio of $\delta$, for
example, $\delta = 1.5$ would mean a woman with a propensity of 0.35
increases to 0.45.
\end{example}



\subsubsection{Modified Treatment Policies}

\begin{example}[Incremental propensity score based on the risk
  ratio]\label{exwen}
  Another type of IPSI 
  begins with a slightly different set-up than \citet{kennedy2019nonparametric}. Instead of relying on the
  treatment mechanism, there is a random draw $\epsilon_t$ from a
  Uniform $(0,1)$ distribution. If the draw is less than some
  $\delta$, then the treatment assignment is the natural value of
  treatment. If not, it is some constant value, for example 0.
  \begin{equation*}
    \text{$\d_t(a_t,\epsilon_t)$} = \begin{cases}
      \text{$a_t$ if $\epsilon_t < \delta$},\\
      \text{$0$ otherwise}
    \end{cases}
  \end{equation*}
  This intervention is said to be based on the risk ratio because $\delta = \pi^\d_t(h_t)/\pi_t(h_t)$, where $\pi^\d_t(h_t)$ is the post-intervention propensity score and $\pi_t(h_t)$ is the propensity score in the observed data generating mechanism. This intervention is motivated by \citet{wen2023intervention}, who propose an intervention where treatment is a random draw from the shifted distribution and results in the same identification formula. \citet{wen2023intervention} argue the risk ratio interpretation may be more intuitive for collaborators, and they demonstrate their intervention in an application studying the effect of PrEP usage increases on sexually transmitted infection rates.

  \end{example}

\subsection{Estimation Algorithms}

\subsubsection{G-Computation Algorithm}

The G-Computation algorithm steps and pseudo-code for a simple example with two time points is shown below. 

\begin{enumerate}
\item Fit a generalized linear model (GLM) for $Y$ conditional on $A_1=a_1$ and
  $H_1=h_1$. Call this $\hat{Q_1}(a_1, h_1)$.
  
  \texttt{Q1\_hat} $\gets$ \texttt{glm}$(\text{outcome} = Y, \text{predictors} = \{H_1, A_1\})$

\item Modify the data set used in step (1) so that the values in the column for $A_1$ are changed to $A^\d_1$. Obtain the predictions for the model $\hat{Q_1}$ using this modified data set. These are pseudo-outcomes
  $\tilde Y_1$.

\texttt{pseudo\_Y1} $\gets$ \texttt{predict}$(\text{fit} =   \texttt{Q1\_hat}, \text{new data} = \{H_1, A_1 = A_1^\d\} )$
  
\item Fit a generalized linear model (GLM) for $\tilde Y_1$ conditional on
  $A_0=a_0$ and $H_0=h_0$. Call this $\hat{Q_0}(a_0, h_0)$.

  \texttt{Q0\_hat} $\gets$ \texttt{glm}$(\text{outcome} = \texttt{pseudo\_Y1}, \text{predictors} = \{H_0, A_0\})$
  
\item  Modify the data set used in step (3) so that the values in the column for $A_0$ are changed to $A^\d_0$. Obtain the predictions for the model $\hat{Q_0}$ using this modified data set. These are pseudo-outcomes
  $\tilde Y_0$.

\texttt{pseudo\_Y0} $\gets$ \texttt{predict}$(\text{fit} =   \texttt{Q1\_hat}, \text{new data} = \{H_0, A_0 = A_0^\d\} )$
  
\item Average $\tilde Y_0$, i.e. compute $\hat\E[\tilde Y_0]$.

\texttt{estimate} $\gets$ \texttt{mean(pseudo\_Y0)}
\end{enumerate}

\subsubsection{Inverse Probability Weighting Algorithm}

The IPW estimator relies on a density ratio $\r_t = \g_t^\d(a_t\mid h_t)/\g_t(a_t\mid h_t)$, where $\g_t^\d(a_t\mid h_t)$ is the density of the intervened exposure, and $\g_t(a_t\mid h_t)$ is the density of the naturally observed exposure. Practically, $\r_t$ can be computed using a clever classification trick proposed in \citet{qin1998inferences, cheng2004semiparametric} and utilized in \citet{diaz_nonparametric_2021}. The IPW algorithm with pseudo-code for a simple example with two time points is as follows. 

\begin{enumerate}
\item Duplicate each row of the data set so that there are $2n$ rows. The first row of a duplicated pair should contain observed values $A_1$, and the second row should be modified so that $A_1 = A_1^\d$. A new column $\Lambda_1$ should be created that is $0$ if $A_1=A_1$ and $1$ if $A_1 = A_1^\d$.

  \texttt{data} $\gets$ \texttt{matrix}$(\{H_1, A_1, \Lambda_1 = 0\} \})$
  
  \texttt{data\_copy} $\gets$ \texttt{matrix}$(\{H_1, A_1 = A_1^\d, \Lambda_1 = 1\})$
  
  \texttt{dr\_data} $\gets$ \texttt{bind\_rows(data, data\_copy)}

\item Fit a logistic regression for $\Lambda_1$ conditional on $A_1=a_1$ and $H_1=h_1$. The odds ratio at a given $a_1$ and $h_1$ is the estimate of the density ratio at time 1, $\hat\r_1$.

  \texttt{r1\_fit} $\gets$ \texttt{glm}$(\text{outcome} = \Lambda_1, \text{predictors} = \{H_1, A_1\}, \text{data} = \texttt{dr\_data})$

  \texttt{r1\_hat} $\gets$  \texttt{predict}$(\text{fit} =   \texttt{r1\_fit}, \text{prediction = odds ratio}, \text{new data = }\texttt{data})$

\item Create a new column in the duplicated data set which contains values of $A_0$ in the first row of a duplicated pair. The second row in a pair should be modified so that $A_0 = A_0^\d$. A new column $\Lambda_0$ should be created that is $0$ if $A_0=A_0$ and $1$ if $A_0 = A_0^\d$.

\begin{figure}[H]
\begin{center}
    \includegraphics[width=8cm]{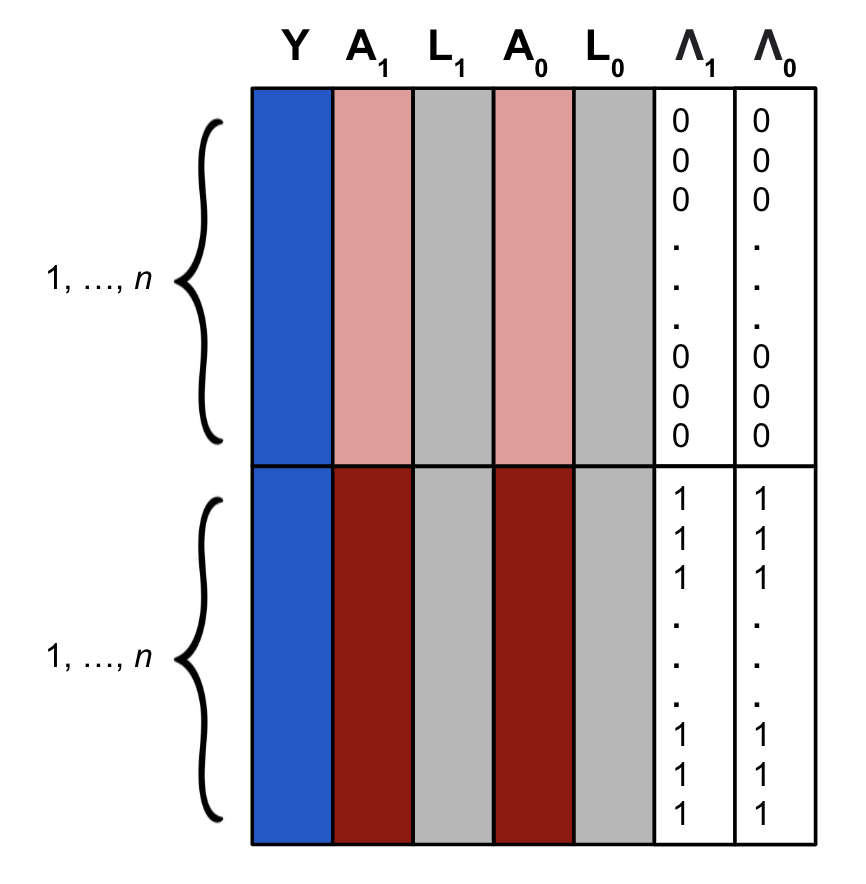}\label{illustration}
\caption{Illustration of the data used for density ratio estimation with the classification trick.}
\end{center}
\end{figure}

\item Fit a logistic regression for $\Lambda_0$ conditional on $A_0=a_0$ and $H_0=h_0$. The odds ratio at a given $a_0$ and $h_0$ is the estimate of the density ratio at time 0, $\hat\r_0$.

 \texttt{r0\_fit} $\gets$ \texttt{glm}$(\text{outcome} = \Lambda_0, \text{predictors} = \{H_0, A_0\}, \text{data} = \texttt{dr\_data})$

  \texttt{r0\_hat} $\gets$  \texttt{predict}$(\text{fit} =   \texttt{r0\_fit}, \text{prediction = odds ratio}, \text{new data = }\texttt{data})$

\item Multiply $\hat\r_1$ and $\hat\r_0$ together. This is the cumulative density ratio. Multiply the cumulative density ratio by $Y$. Compute the average. This is $\hat\E[\tilde{Y}_0]$.

\texttt{estimate} $\gets$ \texttt{mean(r1\_hat * r0\_hat *} $Y)$

\end{enumerate}

\subsubsection{Targeted Maximum Likelihood Estimation (TMLE) Algorithm}

\begin{enumerate}

    \item Perform the same steps as for the IPW algorithm to produce estimates of $\hat\r_1$ and $\hat\r_0$.

    \item Fit a generalized linear model (GLM) for $Y$ conditional on $A_1=a_1$ and $H_1=h_1$. Call this $\hat{Q_1}(a_1, h_1)$.
  
  \texttt{Q1\_hat} $\gets$ \texttt{glm}$(\text{outcome} = Y, \text{predictors} = \{H_1, A_1\})$

  \item Modify the data set used in step (2) so that the values in the column for $A_1$ are changed to $A^\d_1$. Obtain predictions for the model $\hat{Q_1}$ using this modified data set and when $A_0=a_0$ and $H_0=h_0$. 

\texttt{Y1d} $\gets$ \texttt{predict}$(\text{fit} =   \texttt{Q1\_hat}, \text{new data} = \{H_1, A_1 = A_1^\d\} )$ 

\texttt{Y1} $\gets$ \texttt{predict}$(\text{fit} =   \texttt{Q1\_hat}, \text{new data} = \{H_1, A_1\} )$

  \item Fit a logistic regression model for $Y$ with an offset equal to the logit transformation of the expected value of $Y$ under model $\hat{Q_1}(a_1, h_1)$ and weights $\hat{\r_1}$. Call this $\tilde{Q_1}(a_1, h_1)$.

    \texttt{Q1\_tilde} $\gets$ \texttt{glm}($\text{outcome} = Y$, \text{offset} = \texttt{qlogis}(\texttt{Y1}), weights = \texttt{r1\_hat})

  \item Update the predictions \texttt{Y1d} using $\tilde{Q_1}(a_1, h_1)$. These are pseudo-outcomes $\tilde Y_1$. 

    \texttt{pseudo\_Y1} $\gets$ \texttt{plogis(qlogis(Y1d) + coef(Q1\_tilde))}

  \item Repeat steps 2 to 5 replacing $Y$, $A_1$, $H_1$, and $\hat{\r_1}$ with $\tilde Y_1$, $A_0$, $H_0$, and $\hat \r_0 \times \hat\r_1$ to obtain $\tilde{Y}_0$.

  \item Compute $\E[\tilde Y_0]$.
    
    \texttt{estimate} $\gets$ \texttt{mean(pseudo\_Y0)}

\end{enumerate}

\subsubsection{Sequentially Doubly Robust (SDR) Estimation Algorithm}

\begin{enumerate}

    \item Perform steps 1 to 3 as for the TMLE algorithm. 

    \item Using $Y$, $\hat\r_1$, \texttt{Y1}, and \texttt{Y1d} apply the SDR transformation from \citet{diaz_nonparametric_2021} to obtain the pseudo-outcome $\tilde Y_1$.

    \texttt{pseudo\_Y1} $\gets$ \texttt{r1\_hat *} ($Y$ - \texttt{Y1}) \texttt{+ Y1d}

    \item Fit a generalized linear model (GLM) for $\tilde Y_1$ conditional on
  $A_0=a_0$ and $H_0=h_0$. Call this $\hat{Q_0}(a_0, h_0)$.

    \texttt{Q0\_hat} $\gets$ \texttt{glm}$(\text{outcome} = \texttt{pseudo\_Y1}, \text{predictors} = \{H_0, A_0\})$

    \item Modify the data set used in step (3) so that the values in the column for $A_0$ are changed to $a^\d_0$. Obtain predictions for the model $\hat{Q_0}$ using this modified data set and when $A_0=a_0$ and $H_0=h_0$. 

    \texttt{Y0d} $\gets$ \texttt{predict}$(\text{fit} =   \texttt{Q0\_hat}, \text{new data} = \{H_0, A_0 = a_0^\d\} )$ 

    \texttt{Y0} $\gets$ \texttt{predict}$(\text{fit} =   \texttt{Q0\_hat}, \text{new data} = \{H_0, A_0\} )$

    \item Using \texttt{pseudo\_Y1}, \texttt{Y0d}, \texttt{Y0}, \texttt{r1\_hat}, and \texttt{r0\_hat} obtain an estimate of the un-centered efficient influence function, $\hat \phi_1$.

    \texttt{m\_d} $\gets$ \texttt{matrix}(\texttt{Y1d}, $Y$)

    \texttt{m} $\gets$ \texttt{matrix}(\texttt{Y0}, \texttt{Y1})

    \texttt{r} $\gets$ \texttt{matrix}(\texttt{r0\_hat * }\texttt{r1\_hat}, \texttt{r1\_hat})

    \texttt{phi\_hat} $\gets$ \texttt{rowSums}(\texttt{r *} (\texttt{m\_d - m})) \texttt{+ Y0d}

    \item Compute $\E[\hat \phi_1]$

    \texttt{estimate} $\gets$ \texttt{mean}(\texttt{phi\_hat})
    
\end{enumerate}

\subsubsection{Technical requirements for theoretical guarantees of estimators}

 The
  $\sqrt{n}$-consistency and asymptotic normality of the estimators
  proposed in \citet{diaz_nonparametric_2021}, and therefore the
  correctness of the confidence intervals output by the R software
  \emph{lmtp} \citep{nickpaper, lmtppkg}, relies on two important
  technical requirements. 
First, the function $\d$ must not depend on the distribution of the
data. Second, the exposure must be discrete or, if it is continuous,
$\d(\cdot, h_t, \epsilon_t)$ (as a
function of $a_t$) must be piecewise smooth invertible \citep{haneuse2013estimation,
  diaz_nonparametric_2021}. If these requirements are violated,
important theoretical properties of the estimators such as
$\sqrt{n}$-consistent will fail. This means that uncertainty
quantification measures such as confidence intervals, p-values, and
standard errors outputted by the package will be incorrect.

If the first requirement is violated, it may be possible to develop
$\sqrt{n}$-consistent estimators. For example,
\citet{kennedy2019nonparametric} developed non-parametric
$\sqrt{n}$-consistent estimators for the IPSI shown in Example \ref{exkennedy}, in
which $\d$ depends on the treatment mechanism. However, the IPSI in Example \ref{exwen}, which does not
depend on the data distribution, can be estimated with the proposed
estimators. Here it is important to note that the the estimators for
the odds ratio IPSI of \citet{kennedy2019nonparametric} are not doubly
robust (and such estimators possibly cannot be constructed, since $\d$
depends on the propensity score), whereas the estimators of the
risk-ratio IPSI will be doubly robust.
The threshold function as it is in Example \ref{exthreshold} meets these technical requirements because the exposure is discrete, however, if the exposure were instead continuous, it would not
meet the second requirement because a threshold $\d$ is not piecewise smooth
invertible. However, the MTPs in Examples \ref{exsmoking} and 
\ref{exshift} meet
both requirements and are estimable using the proposed estimators.

\subsection{Practical guidance and considerations}

 In practice we recommend using TMLE if the user is not concerned about model misspecification, because the estimates are guaranteed to remain within the observed outcome bounds. However, if model misspecification is a concern, SDR is significantly more robust, although there is a potential for estimates which fall outside the outcome bounds. An effective approach against model misspecification is to use the superlearner algorithm for estimating
the exposure and
outcome mechanisms. Superlearning
combines the predictions of multiple pre-specified statistical learning models via weighting to produce final estimates which are proven to perform as well as possible in large sample sizes 
\citep{vanderLaanPolleyHubbard07}.

To assess positivity violations, we recommend the researcher to visually inspect or examine summary statistics of the density ratios. Extremely high values of density ratios (the quantification of ``high'' being context-dependent) are akin to propensity score values at or near zero, generally indicate positivity violations, and may create unstable estimates. If violations are detected, the researcher may be interested in modifying the portion of the population receiving the intervention, or in cases of continuous exposures, making the intervened exposure level closer to the observed exposure level. Interventions should be designed to avoid both theoretical and practical positivity violations present in the data, the latter of which can potentially be caught in the exploratory data analysis stage. One option if the researcher cannot find a solution by study design to elimitate positivity violations is to truncate the density ratios as a certain threshold, e.g. the 98th quantile. This is akin to truncating a propensity score, and may have potential for biases \citep{leger2022causal}.

Analysts using LMTPs for time-varying or survival outcomes will also note that  discrete outcome intervals are required. The researcher must find a balance in choosing intervals that are both scientifically relevant (e.g. week-long intervals for a critically ill patient population would be too long) and in which there are still a number of outcomes which occur within the chosen interval. If the outcome is rare and the sample size is small, or if the sample size is very large and computationally intensive, the researcher may need to adjust the number of cross-fitting and/or superlearning cross-validation folds.

\subsection{Additional Application Details}

\subsubsection{Motivation}

The motivation for studying a delay in intubation policy for COVID-19 patients is further detailed in \citet{diaz2022causal}'s illustrative application section.

\subsubsection{Methods}

Baseline confounders include age, sex, race, ethnicity, body mass index (BMI), comorbidities
(cerebral vascular event, hypertension, diabetes mellitus, cirrhosis, chronic obstructive pulmonary disease, active
cancer, asthma, interstitial lung disease, chronic kidney disease, immunosuppression, HIV-infection, and home oxygen use), and hospital admission location. Time-dependent confounders include vital signs (heart rate, pulse oximetry percentage,
respiratory rate, temperature, systolic/diastolic blood pressure), laboratory results (blood urea nitrogen (BUN)-
creatinine ratio, creatinine, neutrophils, lymphocytes, platelets, bilirubin, blood glucose, D-dimers, C-reactive
protein, activated partial thromboplastin time, prothrombin time, arterial partial pressures of oxygen and carbon
dioxide), and concurrent pharmaceutical treatments. Concurrent treatments included vasopressors, diuretics, Angiotensin-converting enzyme
(ACE) and Angiotensin receptor blockers (ARBs), hydroxychloroquine, and tocilizumab.

In the case of multiple time-varying
confounders measured in one day, the clinically worst value was
used.
Although data was relatively complete due to manual abstraction
efforts, there were instances of patients missing laboratory
results at one or multiple time points. A combination of last observation carried forward and an
indicator for missing values was used to handle this informative missingness within the treatment and outcome regression models \citep{burton2004missing}. 

The Superlearner ensemble algorithm utilized 5-fold cross-validation and candidate libraries included generalized linear models, multivariate adaptive regression splines \cite{earth}, random forests \cite{ranger}, and
extreme gradient boosted trees \cite{xgboost} An assumption was made that the time-varying
covariates from the previous two days (i.e. lag of 2 days) was
sufficient to capture the mechanisms to reflect the collection of
laboratory results at minimum 48-hour intervals. A 5-fold
cross-fitting component was implemented for the final estimator to
prevent variation in a certain sample split from biasing the final
estimate \citep{zivich2021machine}. All analyses were conducted in R Version 4.1.2 with the packages tidyverse \cite{tidyverse} for data cleaning and plotting, and lmtp \cite{lmtppkg} and SuperLearner \cite{superlearner} for estimation.

\subsubsection{Discussion}

 There are limitations specific to our illustrative example. While the estimated effect of a less aggressive intubation strategy can help to understand an underlying biological or mechanistic process, it may not provide clinical guidance if the treatment strategy changes over time. The estimates themselves depend on the natural value of treatment, and this is dependent on the state of clinical practice during the study time frame (Spring 2020). 
In addition, we cannot rule out unmeasured confounding in the
exposure, outcome, and loss to follow-up mechanisms. We also cannot be sure whether the informative right censoring is correctly specified, since patients were lost to follow-up for different reasons (e.g. discharge to home vs. assisted living). 

\subsection{Further reading}

Researchers intending to learn more about LMTPs may find the original LMTP methods research by \citet{diaz_nonparametric_2021} useful, as well the extension to competing risks (with a similar application as this tutorial) in \citet{diaz2022causal}. The R software \textit{lmtp} is explained in \citet{nickpaper}. 

The LMTP methodology is based upon decades of work in epidemiology, biostatistics, and related fields. Interested readers should see related methods research and applications such as \citet{robins2004effects, taubman2009intervening, munoz2012population, haneuse2013estimation, YoungHernanRobins2014, richardson2013single}. 

\pagebreak


\bibliography{refs}